\documentclass[10pt]{iopart}

\usepackage{iopams}  

\usepackage{float,color,multicol,xspace,subfigure,dcolumn,tabularx,bm}
\usepackage{xspace}

\usepackage[square,comma,numbers,sort]{natbib}
\usepackage{graphicx}

\usepackage{type1cm,eso-pic}
\usepackage{bm}

\newcommand{\AlGaN}{$\rm{Al}_x\rm{Ga}_{1-x}\rm{N}$\xspace}
\newcommand{\AlInN}{$\rm{Al}_x\rm{In}_{1-x}\rm{N}$\xspace}
\newcommand{\GaInN}{$\rm{Ga}_x\rm{In}_{1-x}\rm{N}$\xspace}
\newcommand{\AlGaInN}{$\rm{Al_x}\rm{Ga}_{y}\rm{In}_{1-x-y}\rm{N}$\xspace}

\newcommand{\todo}[1]{}
\newcommand{\myStrut}{\rule{0pt}{4ex}}

\begin{document}

\title[Computational study of structural and elastic properties of AlGaInN]
{Computational study of structural and elastic properties of 
  random \AlGaInN alloys}

\author{M \L{}opuszy\'nski }

\address{ 
Interdisciplinary Centre for Mathematical and Computational Modelling, 
University of Warsaw, Pawi{\'n}skiego 5A, 02-106 Warsaw, Poland
}
\ead{m.lopuszynski@icm.edu.pl}

\author{J A Majewski}
\address{
Institute of Theoretical Physics, Faculty of Physics,
University of Warsaw, Ho{\.z}a 69, 00-681 Warsaw, Poland
}

\begin{abstract}
In this work we present a detailed computational study of structural and
elastic properties of cubic \AlGaInN alloys in the framework of Keating valence
force field model, for which we perform accurate parametrization based on state
of the art DFT calculations.  When analyzing structural properties, we focus on
concentration dependence of lattice constant, as well as on the distribution of
the nearest and the next nearest neighbour distances.  Where possible, we
compare our results with experiment  and calculations performed within other
computational schemes.  We also present a detailed study of elastic constants
for \AlGaInN alloy over the whole concentration range. Moreover, we include 
there accurate quadratic parametrization for the dependence of the alloy
elastic constants on the composition.  Finally, we examine the sensitivity of
obtained results to computational procedures commonly employed in the Keating
model for studies of alloys. 
\end{abstract}


\section{Introduction}
Recently AlN, GaN and InN trigger a lot of interest in optoelectronics,
mostly because of their electronic structure, which makes them very promising
materials for application in blue/green and UV active devices.  Emitters and
detectors operating in these spectral range can be used in many important areas
such as optical data storage, biosensors, multimedia etc.  Nitride alloys allow
for continuous tuning of physical quantities such as band gap, lattice
parameters, mobility etc., to reach the suitable values for desired
applications.

Despite both theoretical and experimental efforts, many basic properties of
nitride alloys are not yet sufficiently well understood.  In this work, we
focus on calculations of structural and elastic properties of quaternary
\AlGaInN alloys, since they offer the largest possibility of tuning.  In
particular, we examine the morphology of alloys concentrating on atomic
distance distribution between the nearest and the next nearest neighbours.
They have important influence on the electronic structure of alloys (see e.g.
\cite{Gorczyca2009,Gorczyca2009a}). This knowledge is crucial for application
purposes. In the case of simpler ternary alloys AlGaN \cite{Miyano1997,Yu1999},
AlInN \cite{Katsikini2008}, GaInN
\cite{Jeffs1998,Katsikini2003,Kachkanov2006,Katsikini2008,Miyanaga2007}, the
bond lengths distribution has been  obtained by extended X-ray absorption fine
structure spectroscopy (EXAFS), however, for quaternary \AlGaInN there are no
available experimental data.  As far as elastic properties of alloys are
concerned, their correct description is also very important issue in modelling
lasing from quantum wells, e.g. within $\bm{k} \cdot \bm{p} $ or similar
continuous models. The nonlinear effects in elasticity of nitrides and their
influence on properties of devices have recently attracted considerable
attention {\cite{Lepkowski2004,Lepkowski2005,Lopuszynski2007,Lepkowski2007}.
However, there is not much known about the detailed dependence of elastic
properties on composition.  In this paper, we present a computational study of
the second-order elastic constants, $c_{ij}$, for  \AlGaInN quaternary alloys
within the whole concentration range.  This could provide very useful insight
from viewpoint of device modelling.

In the present study of quaternary nitride alloys, our main computational tool
is valence force field approach (VFF) developed by Keating \cite{Keating1966}.
Even though it  was developed over forty years ago, it is still important
ingredient of multiscale models, particularly where large number of alloy
configurations needs to be handled.  For nitrides, the Keating VFF model has
been recently used to examine a plethora of physical phenomena, such as phonon
spectra in bulk and in superlattices \cite{Zi1996,Zi1996a,Wei1997}, structural
properties of ternary bulk alloys
\cite{Bellaiche1997,Mattila1999,Saito1999,Grosse2001, Chen2008} and their
nanowires \cite{Xiang2008}, stability  of different alloy phases
\cite{Chen2008} and also in numerous studies of thermodynamics of ternary
\cite{Ho1996, Takayama2000, Takayama2001, Adhikari2004, Karpov2004, Biswas2008}
and quaternary \cite{Adhikari2004a} nitride alloys.  Generally, the Keating VFF
model is also a method of choice, where the atomic positions are needed as
external input for electronic structure modelling. This is the case for methods
atomistic in nature, but not entirely based on first principles, such as
semiempirical tight-binding or empirical pseudopotential methods, that are
commonly used for studies of low dimensional semiconductor structures.
Therefore, to contribute to further development and validation of the model, we
also pay attention to practical aspects of VFF usage.  We compare the
distribution of the nearest neighbour and next nearest neighbour distances
resulting from Keating VFF model with those obtained from accurate quantum
mechanical formalism, which is a good test of VFF model reliability.  We also
examine the influence of so called mixing rule used to obtain VFF parameters 
for alloys, which shows how strongly this could influence prediction of 
this model.

The paper is organized as follows. In section \ref{Sec:KeatingModelParam} we
briefly recall basic facts about Keating model and present the employed set of
parameters.  Section \ref{Sec:StrucProps} provides detailed overview of
structural properties for quaternary nitride alloys, it includes lattice
constants as well as the distribution of the nearest neighbour and the next
nearest neighbour distances.  In this part we also compare the results of
Keating VFF with DFT findings.  In section \ref{Sec:ElasticConstants}, the VFF
results for elastic constants of \AlGaInN alloys in the whole concentration
range are presented. Section \ref{Sec:MixingInfluence} deals with the
computational procedures, so-called mixing rule used to obtain VFF parameters of
alloys and the effect of finite supercell size.  Finally, the paper is
summarized in section \ref{Sec:Summary}.

\section{Keating valence force field model and its parametrization
\label{Sec:KeatingModelParam}}

Keating \cite{Keating1966}, on the basis of general symmetry considerations, 
derived potential energy model for zinc-blende type crystals
in the following form
\begin{eqnarray}
 V(\bm{r}_{1}, \bm{r}_{2}, \dots) &=&
 \sum_{i}  \sum_{j \in NN(i)}
     \frac{3\alpha_{ij}}{16 d_{ij}^2} 
	     \left ( \bm{r}_{ij}^2 - d_{ij}^2 \right)^2 + \\
&&
   \sum_{i}  
   \sum_{ { j,k \, \in NN(i)} \atop {j \neq k }}
     \frac{3 \beta_{ijk}}{16 d_{ij} d_{ik}} 
	   \left( \bm{r}_{ij} \cdot \bm{r}_{ik} - 
	           d_{ij} d_{ik} \cos{\theta_0} \right)^2. \nonumber
\end{eqnarray}
Here, $\bm{r}_{ij}=\bm{r}_i-\bm{r}_j$, where $\bm{r}_i$, $\bm{r}_j$ denote the
position of i-th and j-th atoms respectively, $d_{ij}$ denotes equilibrium
distance between atom i and j, $NN(i)$ represents the set of four nearest
neighbours of i-th atom, $\alpha_{ij}$ and $\beta_{ijk}$ stand for force
constants. 

In the studies applying VFF to nitrides, very often a set of parameters
proposed by Kim \etal \cite{Kim1996} is used. However, during the last years
theoretical suggestions appeared, that propose refined procedure for
determination of the force constants in VFF model \cite{Grosse2001}. On the
other hand, the overall improvement of accuracy of the first-principles
computational methods has been also achieved. Therefore, we have decided to
recalculate the force constants and bring them to the state-of-the-art. 

The parametrization of the Keating VFF model starts from determination of the
force constants $\alpha$ and $\beta$ for bulk zinc-blende compounds. This can
be done knowing the elastic constants of the bulks. As shown by Keating
\cite{Keating1966}, the $\alpha$ and $\beta$ are given by analytic formula as a
function of the elastic constants $c_{11}$ and $c_{12}$. In the Keating model,
the third elastic constant, i.e., $c_{44}$, is related to $c_{11}$ and $c_{12}$
as follows
\begin{equation}
\label{Eq:KeatingRelation}
\frac{2 c_{44} (c_{11}+c_{12})}{(c_{11}-c_{12})(c_{11}+3c_{12})}=1.
\end{equation}
Therefore, in the standard procedure to determine the $\alpha$ and $\beta$,
$c_{44}$ is not taken into account. However, the equation
(\ref{Eq:KeatingRelation}) for elastic constants resulting from Keating model
is not very well satisfied for nitrides.  To improve this approach, Grosse and
Neugebauer \cite{Grosse2001} proposed an alternative method.  They suggested to
determine $\alpha$ and $\beta$ by the least-squares fit to all three elastic
constants. It turns out that such fitting approach ensures more uniform
spreading of error and generally leads to better results.  Therefore, we follow
this approach in our work.  Since the values of elastic constants for
zinc-blende nitrides have not been measured so far, we relay on theoretical
predictions from the DFT based calculations. In our study, we take arithmetic
average of elastic constants obtained in the calculations within the 
DFT LDA and DFT GGA approximations to determine the force constants $\alpha$ and
$\beta$. The elastic constants obtained through the averaging over these two
theoretical schemes should be closer in value to experimental ones. It follows
from the DFT LDA and DFT GGA tendencies to, respectively, overestimate and
underestimate the stiffness of a material. The values of $c_{ij}$ calculated
within GGA formalism have been taken from our previous work
\cite{Lopuszynski2007}, and the $c_{ij}$ values have been calculated
within DFT LDA approximation using VASP package
\cite{Kresse1994,Kresse1996,Kresse1996a} and the projector augmented wave
method \cite{Kresse1999}. The energy cutoff was set to 800 eV and the
11$\times$11$\times$11 Monkhorst-Pack mesh \cite{Monkhorst1976} has been
employed for the Brillouin zone integrals.  The values of elastic constants
used for the fitting procedure are summarized in the table
\ref{Tab:ElasticConstants}, whereas the  final set of employed parameters is
shown in the table \ref{Tab:KeatingParameters}.

\begin{table}[ht!]
\caption{\label{Tab:ElasticConstants}
Elastic constants used to parametrize Keating model.}
\vspace{0.2cm}
\begin{indented}
\item[] \begin{tabular}{l|c|c|c|l}
 & $c_{11}$ & $c_{12}$ & $c_{44}$ & \\
\hline
\rule{0pt}{3ex}
AlN  & 301 & 167 & 188  & LDA \\
     & 282 & 149 & 179  & GGA, see \cite{Lopuszynski2007} \\
GaN  & 288 & 159 & 161  & LDA \\
     & 252 & 129 & 147  & GGA, see \cite{Lopuszynski2007} \\    
InN  & 184 & 127 &  84  & LDA \\
     & 159 & 102 &  78  & GGA, see \cite{Lopuszynski2007}
\end{tabular}
\end{indented}
\end{table}

Now we are in the position to determine parameters $\alpha_{ij}$ and
$\beta_{ijk}$ for calculation of \AlGaInN quaternary alloys.  The values of the
force constants $\alpha_{ij}$ are directly taken as force constants $\alpha$ of
binary compound consisting of atomic species $i$ and $j$. In the case of
$\beta_{ijk}$ three types of atoms can be involved, and obviously the
parametrization, carried out for binary materials cannot determine these
constants.  In such cases we employ arithmetic mixing rule, i.e., take
arithmetic average of suitable binary parameters, as it has been applied in
many previous works, e.g. \cite{Mattila1999,Takayama2000,Takayama2001}.
Specifically, this means that $\beta_{\rm{Ga},{N},{In}} =
(\beta_{\rm{Ga},{N},{Ga}} + \beta_{\rm{In},{N},{In}} )/2$, and so on.  However,
there is also possibility to use geometric average instead (see e.g. Schabel
and Martins \cite{Schabel1991} or Saito and coworkers \cite{Saito1999}).  We
investigate the role of employed mixing rule later on (see section
\ref{Sec:MixingInfluence} for details). 

\begin{table}[ht!]
\caption{\label{Tab:KeatingParameters}
Set of parameters for Keating VFF model employed in the present study.}
\vspace{0.2cm}
\begin{indented}
\item[] \begin{tabular}{l|c|c|c}
& d  $[\rm{\AA}] $ & 
$\alpha$ $[\rm{N}/\rm{m}]$  & 
$\beta$  $[\rm{N}/\rm{m}]$ \\
\hline
\rule{0pt}{3ex}
 AlN & 1.894 & 79.91 & 19.73 \\
 GaN & 1.950 & 76.25 & 17.80 \\
 InN & 2.165 & 62.07 & 9.68 \\
\end{tabular}
\end{indented}
\end{table}

\section{Structural properties from VFF model and their comparison with
DFT calculations \label{Sec:StrucProps} }

In this section, we analyze composition dependence of structural properties and
geometry of \AlGaInN alloys (i.e., lattice constants, the nearest and next
nearest neighbour distances).  The distribution of bond lengths in an alloy can
be also extracted from EXAFS experiments, which is a useful crosscheck for
the theory.
The local geometry of alloy is an important issue, since it
influences the electronic structure, see e.g. recent calculations of
Gorczyca \etal \cite{Gorczyca2009,Gorczyca2009a} for nitride alloys.  The
Keating VFF model is particularly suitable for the task of establishing local
environment of random alloys. It enables calculations with large supercells,
which in turn guarantee reasonably good sampling and randomness of alloys.
There is fairly long history of employing VFF models to determine alloy
geometry for various types of materials.  For example, Cai and Thorpe studied
relaxation patterns in general ternary $A_{1-x}B_x C$ \cite{Cai1992} and
quaternary $A_{1-x}B_x C_{1-y}D_y$ alloys \cite{Cai1992a} using Kirkwood model
very similar to Keating VFF approach.  Schabel and Martins \cite{Schabel1991}
gave detailed overview of structure for very broad range of semiconducting
alloys within Keating VFF. Their work, however, does not include nitrides.  The
structure of ternary nitride alloys has been also studied by many other authors
\cite{Bellaiche1997,Mattila1999, Saito1999}.  In this work, we present the
first VFF study of structure for quaternary nitride alloys. We compare our
results for bond length distribution with recent calculations of  Marques \etal
\cite{Marques2006} carried out within generalized quasichemical approximation
(GQCA).  Where possible we also compare our findings with experimental data,
obtained using EXAFS approach. To shed some light on accuracy of Keating VFF,
we also compare the structural properties of this force field model with
accurate DFT calculations.  The latter treatment allows only for calculations
with moderate sizes of supercells.  On the other hand DFT description of
interactions between atoms is otherwise very complete. Therefore, such
comparison can reveal potential weak points of fully local classical
description.  Since VFF model is widely used to provide geometry to
semi-empirical calculations of electronic structure handling large systems
(e.g. tight-binding or empirical pseudopotential schemes), the question of its
accuracy is vivid.  Obviously, the quality of such multiscale approach depends
on accuracy of input structures, however, to the best of our knowledge, no
detailed comparison of VFF with more accurate models (e.g. DFT) has been so far
presented in the literature.
 
\subsection{Computational details}

For VFF simulations, we used $18\times18\times18$ zinc-blende cubic supercells
containing 46656 atoms with random distribution of cations.  We optimized both
the atomic positions and the lattice constant of the cubic cell.  For the DFT
simulations, we employed VASP package \cite{Kresse1994,Kresse1996,Kresse1996a}.
We used local density approximation for exchange and correlation functional
according to Ceperley and Alder \cite{Ceperley1980}.  The projector augmented
wave method was used in its variant available in the VASP code
\cite{Kresse1999}.  The calculations were performed for $3\times3\times3$
zinc-blende cubic cells containing 216 atoms. Also in this case cation
distribution among sites was random. The energy cutoff was set to 550 eV.  For
Brillouin zone sampling, only $\Gamma$ point was used due to the large supercell
size.

\subsection{Lattice constant}
In the early 20th century Vegard noticed that lattice constant for alloys can
be calculated using linear interpolation between lattice constants of
constituents \cite{Vegard1921}.  Nitride alloys follow this so-called Vegard's
law, which has been confirmed  by a series of experimental findings, see e.g.
recent experiment for GaInN nanowires, where the linear dependence of lattice
constant was observed in the whole concentration range \cite{Kuykendall2007}. 

The results of our simulations for lattice constant are presented in figure
\ref{Fig:LatticeConstants}. One can easily observe that DFT LDA formalism
reproduces very well the linear Vegard-like behaviour, even though the LDA
approximation is well known to systematically underestimate the values of
lattice constants.  On the other hand the Keating VFF predicts some bowing in
the dependence of lattice constant on concentration.  This is in accordance
with the work of Thorpe and coworkers \cite{Thorpe1991}, who analyzed the
Vegard's law for a simple VFF-like model.  In their case they showed that
Vegard's law is obtained when force constants disorder is neglected.
Therefore, since in our calculations force constant disorder is included (i.e.,
$\alpha$ and $\beta$ depend on the types of atoms considered), it is possible
to expect deviations from linearity.  This is a feature of the Keating VFF
method, however, as can be seen from figure \ref{Fig:LatticeConstants}, its
magnitude is not very large.

\begin{figure}[!ht]
	\includegraphics[height=0.5\textwidth,angle=270]
       {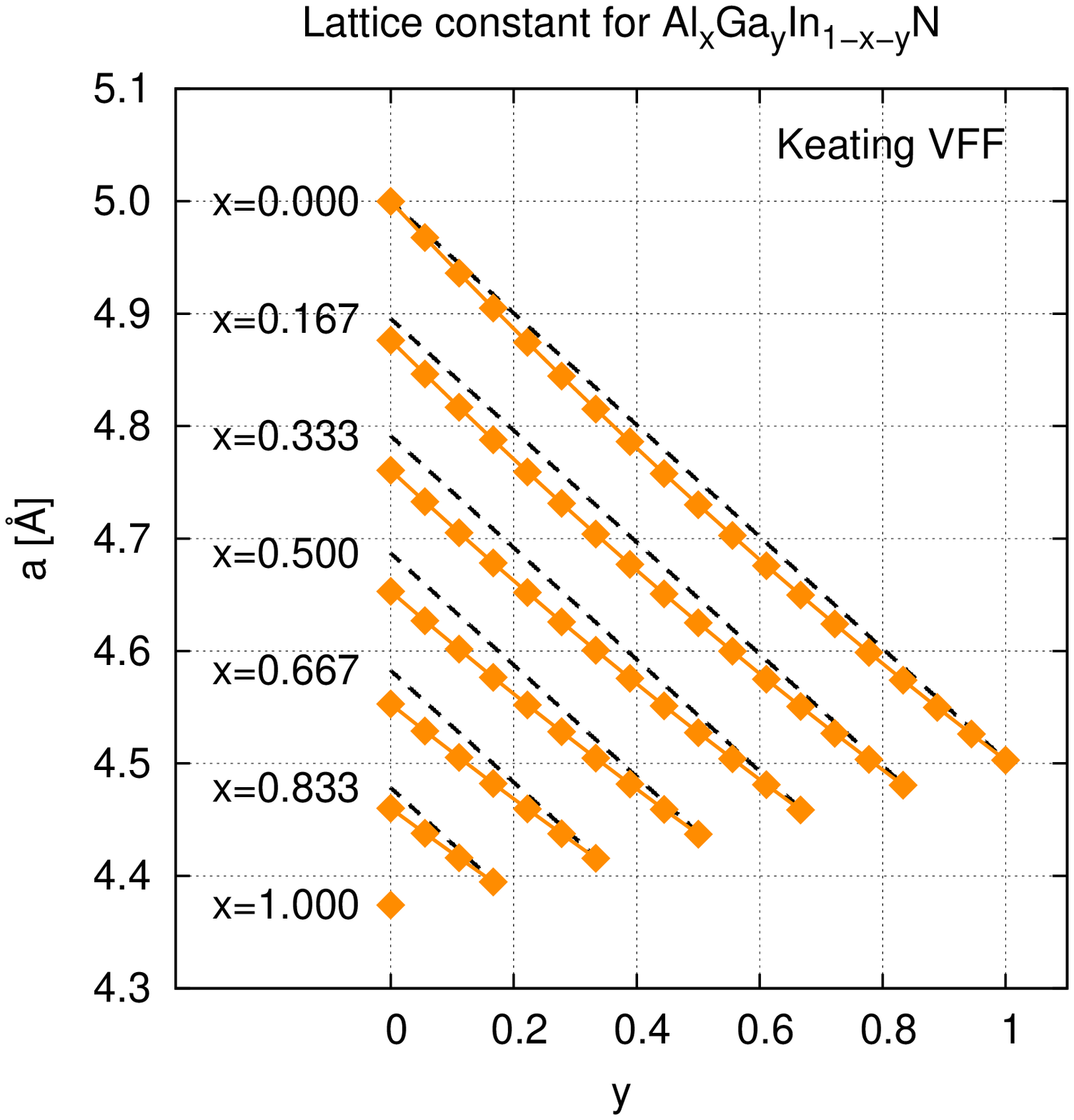}
	\includegraphics[height=0.5\textwidth,angle=270]
       {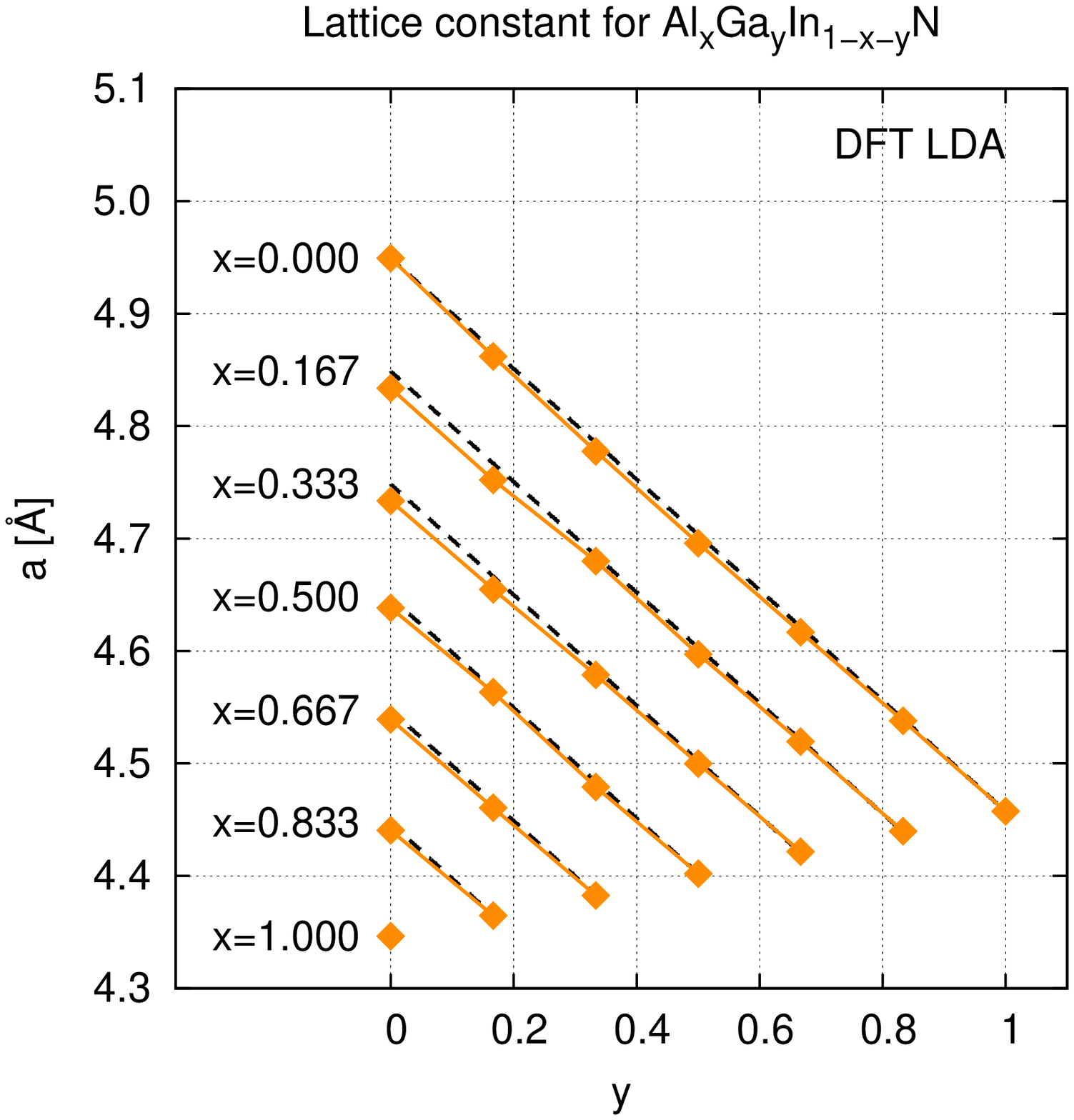}
	\caption{
	\label{Fig:LatticeConstants}
   Comparison of lattice constants dependence on the alloy composition for
   \AlGaInN obtained using Keating VFF (left panel) and DFT LDA (right panel).
   Points correspond to results of calculations, solid lines are only to guide
   the eye. Dashed lines denote prediction of Vegard's law. Note that for
   convenient comparison the scale on both graphs is the same.  }
\end{figure}

\subsection{Nearest neighbours distance}

Results of calculations of the nearest neighbour distance dependence on
concentration of alloy constituents for \AlGaInN  are presented in figure
\ref{Fig:BondLengths}.  It is well known that even though lattice constants
obey Vegard's law, the individual bond lengths do not follow this simple rule.
The dependence of bond lengths on concentration is usually also linear,
however, bond lengths remain much closer to their original bond length in bulk
binary material, rather then to the average bond length predicted by
Vegard-like law (see e.g. \cite{Martins1984,Chen1995}).  This linear dependence
of the bond lengths in the alloy on the concentration of Al and Ga cations,
$x$ and $y$ respectively, can be described by the linear form 
\begin{equation}
\label{Eq:LinearFit} d(x,y)=d_0+Ax+By.  
\end{equation} 
The coefficients $A$ and $B$ in equation (\ref{Eq:LinearFit}) have been
determined by fitting the bond lengths obtained from both Keating and LDA
schemes. These linear fits are represented by dashed lines in figure
\ref{Fig:BondLengths}.  The small values of coefficients A and B (see table
\ref{Tab:BondLengthsFits} gathering results of fitting procedure for all types
of bonds) allow even for rough approximation of $d(x,y)$ by $d_0$, just 
confirming the statement about bond lengths made above.

On the basis of the presented plots (figure \ref{Fig:BondLengths}) for bond
lengths in  \AlGaInN  alloy, one can compare findings of Keating VFF model with
DFT LDA approach. Results of both computational methods  can be reasonably well
described by linear dependence of bond lengths on cation concentrations given
by equation (\ref{Eq:LinearFit}).  Naturally, since the DFT LDA calculations
have been performed for much smaller cell sizes (histograms were generated on
the basis of two 216 atoms supercell calculations) the data have larger
statistical error than the Keating VFF computations. This effect is
particularly pronounced in these parts of the graphs where low concentration of
considered cation is present in the sample. Second evident observation is that
DFT LDA bond lengths are systematically shifted towards the lower values than
Keating VFF. This is related to well known LDA flaw to underestimate the
lattice constants.  It is also worth noticing that results of previous
computations by Marques \etal \cite{Marques2006}, which have been carried out
within generalized quasichemical approximation (GQCA), are in particularly
good agreement with findings of Keating VFF model (see table
\ref{Tab:BondLengthsFits} for details).

Another interesting aspect is to compare the probability distribution profiles
predicted by both DFT LDA and Keating VFF models.  Two sample histograms
generated by both models are depicted in figure \ref{Fig:BondHistograms}. One
can see that both schemes lead to results that are in satisfactory agreement
with each other.  It is again visible that DFT LDA systematically
underestimates the bond lengths, so the peaks corresponding to each bond type
are shifted towards lower values. One can also notice that Keating VFF model
predicts larger peak widths than the DFT LDA approach does.  Particularly
interesting are cases of high In concentration. There, one can observe rather
large lattice deformations caused mostly by the considerable lattice mismatch
between InN and GaN or AlN. This can be considered as a test of the Keating VFF
which has been parametrized on the basis of elastic constants $c_{ij}$, i.e.,
taking into account effects of second-order in deformation strain, which, in
principle, describe the material behaviour in the regime of small deformations.
However, also for these cases, the agreement is reasonable, in spite of
neglecting higher order contributions to the elastic energy (see e.g. results
for Al$_{0.17}$Ga$_{0.17}$In$_{0.66}$N presented in figure
\ref{Fig:BondHistograms}).

\begin{table}[ht!]
\caption{\label{Tab:BondLengthsFits}
         Results of the linear fits to average nearest neighbours distances
         for \AlGaInN quaternary alloys.}
\vspace{0.2cm}
\begin{indented}
\item[] \begin{tabular}{l|c|c|c|c}
\multicolumn{1}{c|}{Bond} & Model & Fit Function & Error [\%]  
      & Ref.\\
\hline
Al-N  
      & VFF & $ 1.9496 -0.05496\,x -0.03893\,y$ & 0.10 
	  & This work \\
      & DFT LDA & $1.9279 - 0.04419\,x - 0.04314\,y$ & 0.26 
      & This work \\ 
      & GQCA    & $1.9435 - 0.05090\,x - 0.05455\,y$ & 0.19 
      & \cite{Marques2006} \\
\hline	  
Ga-N  
      & VFF     & $1.9959 - 0.06322\,x - 0.04573\,y$ & 0.10 
	  & This work  \\
      & DFT LDA & $1.9762 - 0.04367\,x - 0.04470\,y$ & 0.27
      & This work \\
      & GQCA    & $1.9919 - 0.05985\,x - 0.06050\,y$ & 0.35 
	  & \cite{Marques2006} \\
\hline
In-N  
      & VFF     & $	2.1676 -0.09899\,x -0.07485\,y$  & 0.12    
      & This work\\ 
      & DFT LDA & $2.1461 - 0.05276\,x - 0.05088\,y$ & 0.21
      & This work \\
      & GQCA    & $2.1573 - 0.08292\,x - 0.07570\,y$ & 0.55 & 
	  \cite{Marques2006} 
\end{tabular}
\end{indented}
\end{table}

\begin{figure}[p]
	\includegraphics[height=0.5\textwidth,angle=270]
        {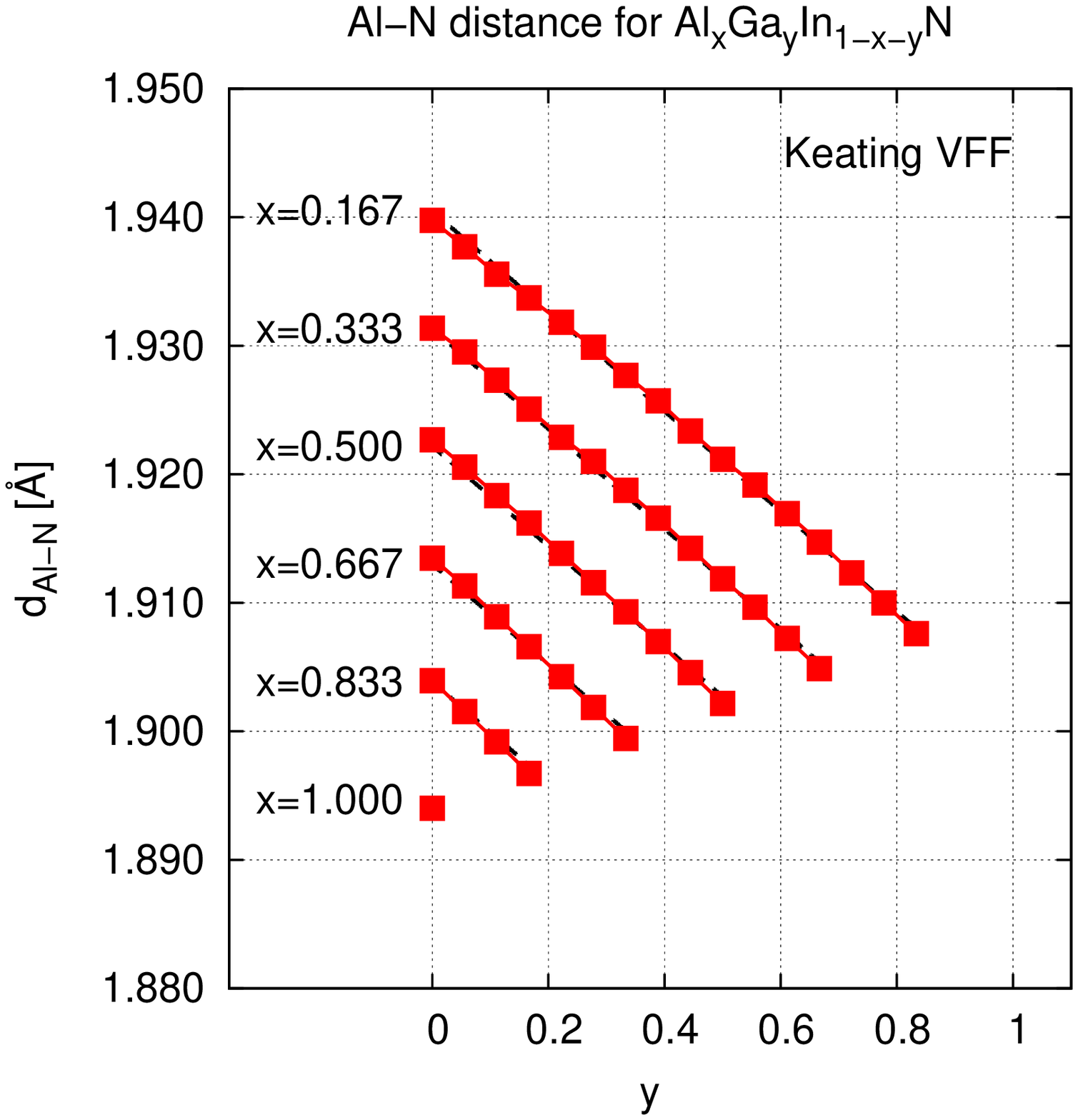}
    \includegraphics[height=0.5\textwidth,angle=270]
        {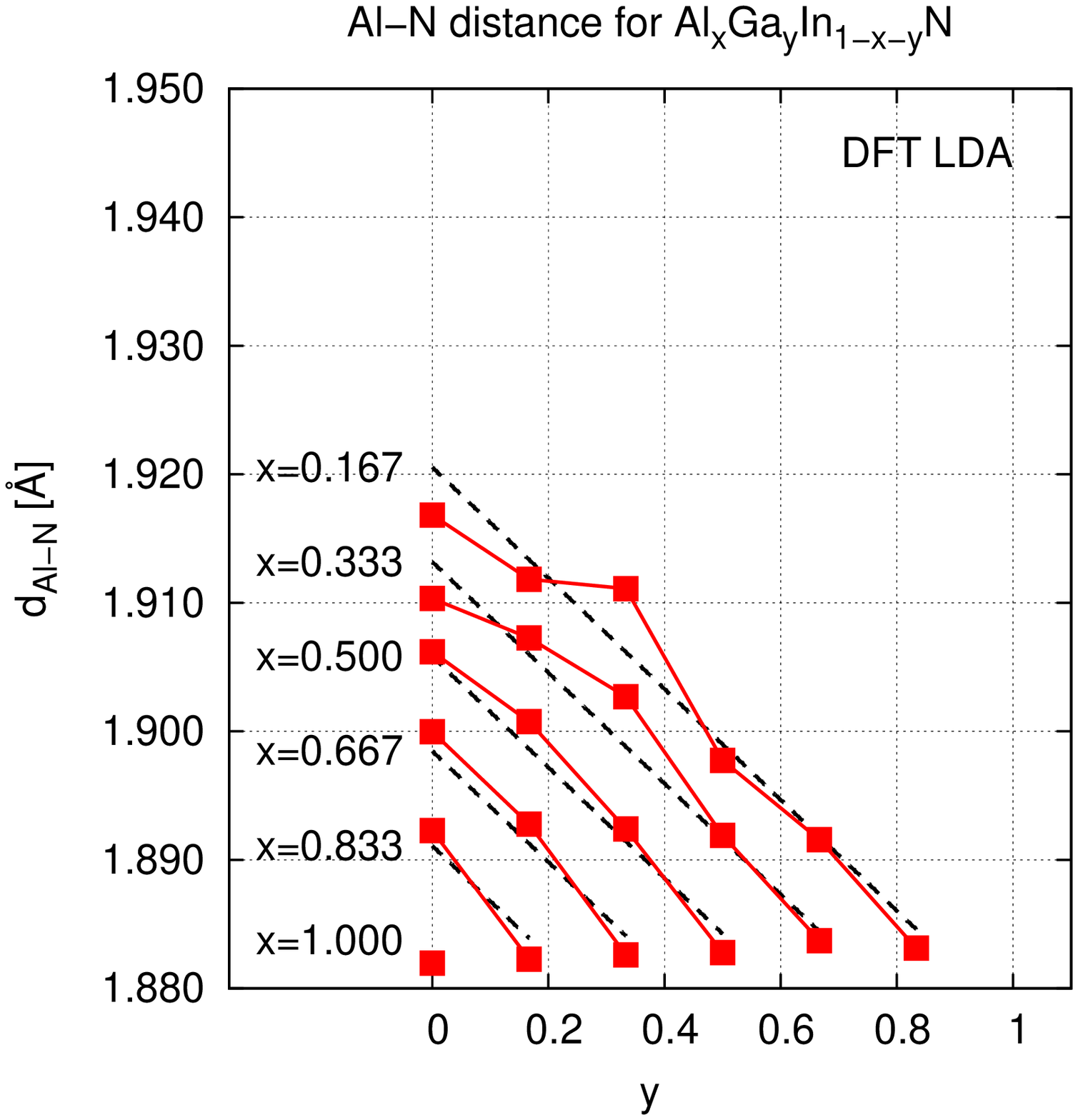}
	\includegraphics[height=0.5\textwidth,angle=270]
        {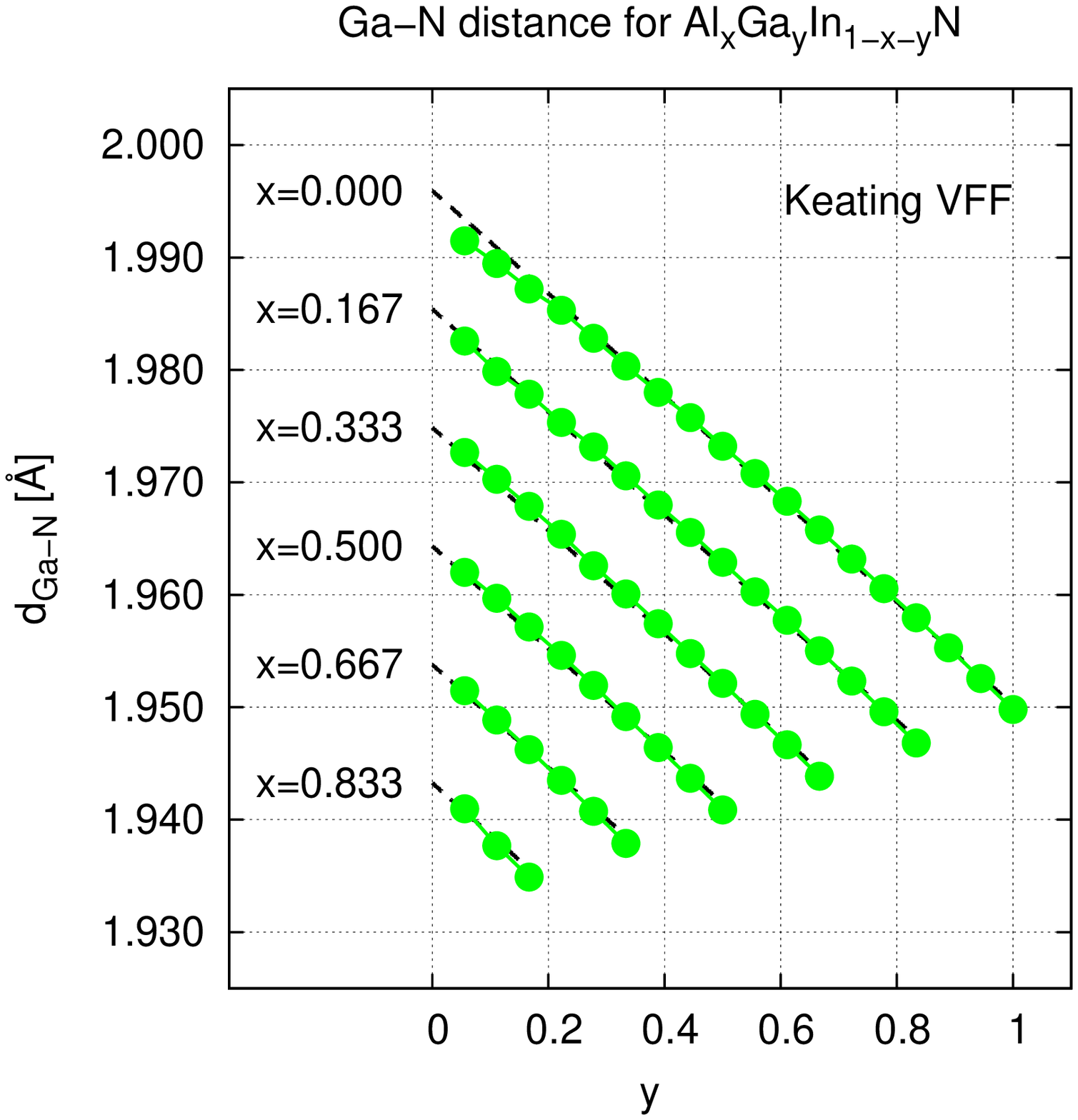}
	\includegraphics[height=0.5\textwidth,angle=270]
        {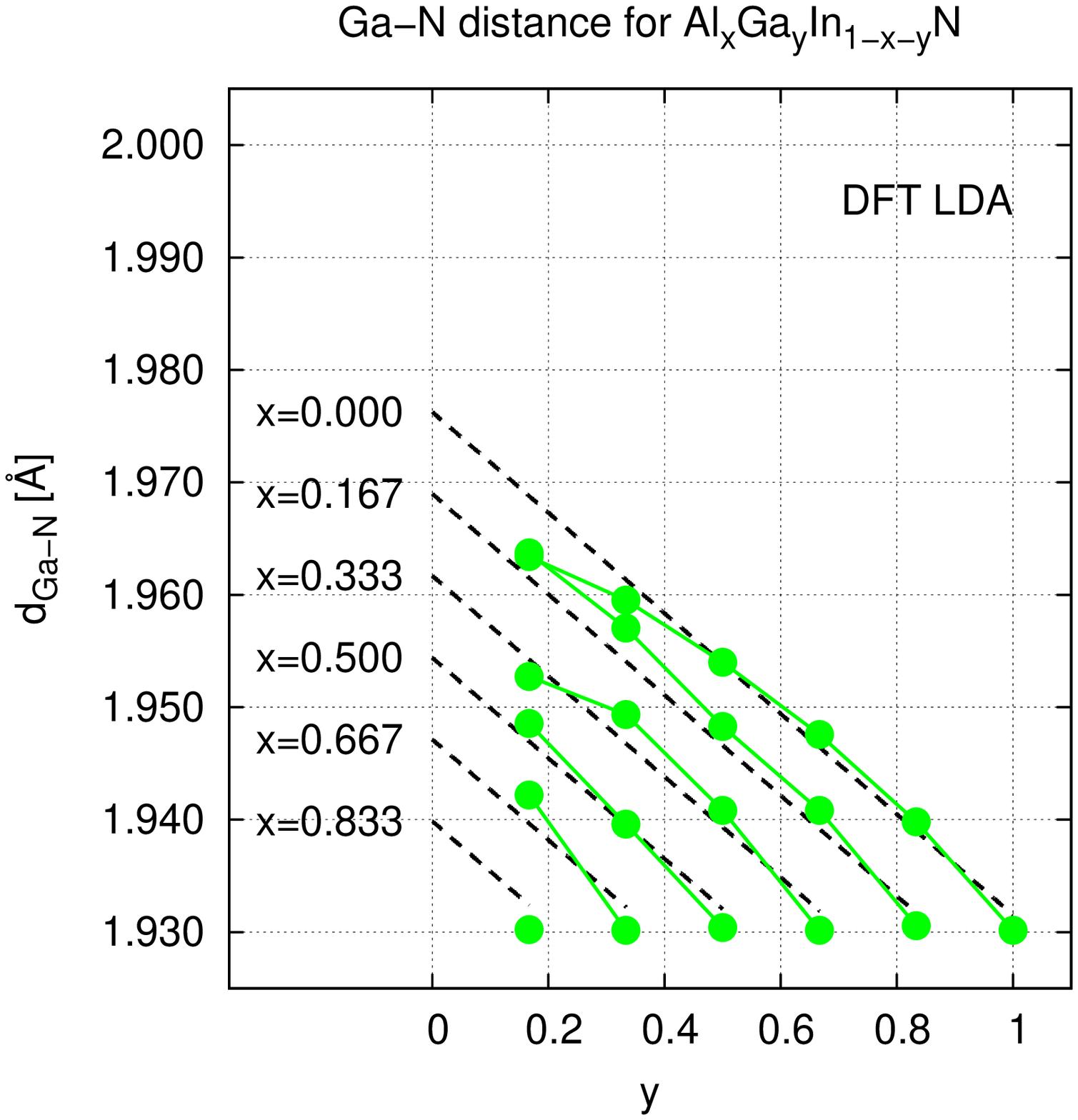}
	\includegraphics[height=0.5\textwidth,angle=270]
        {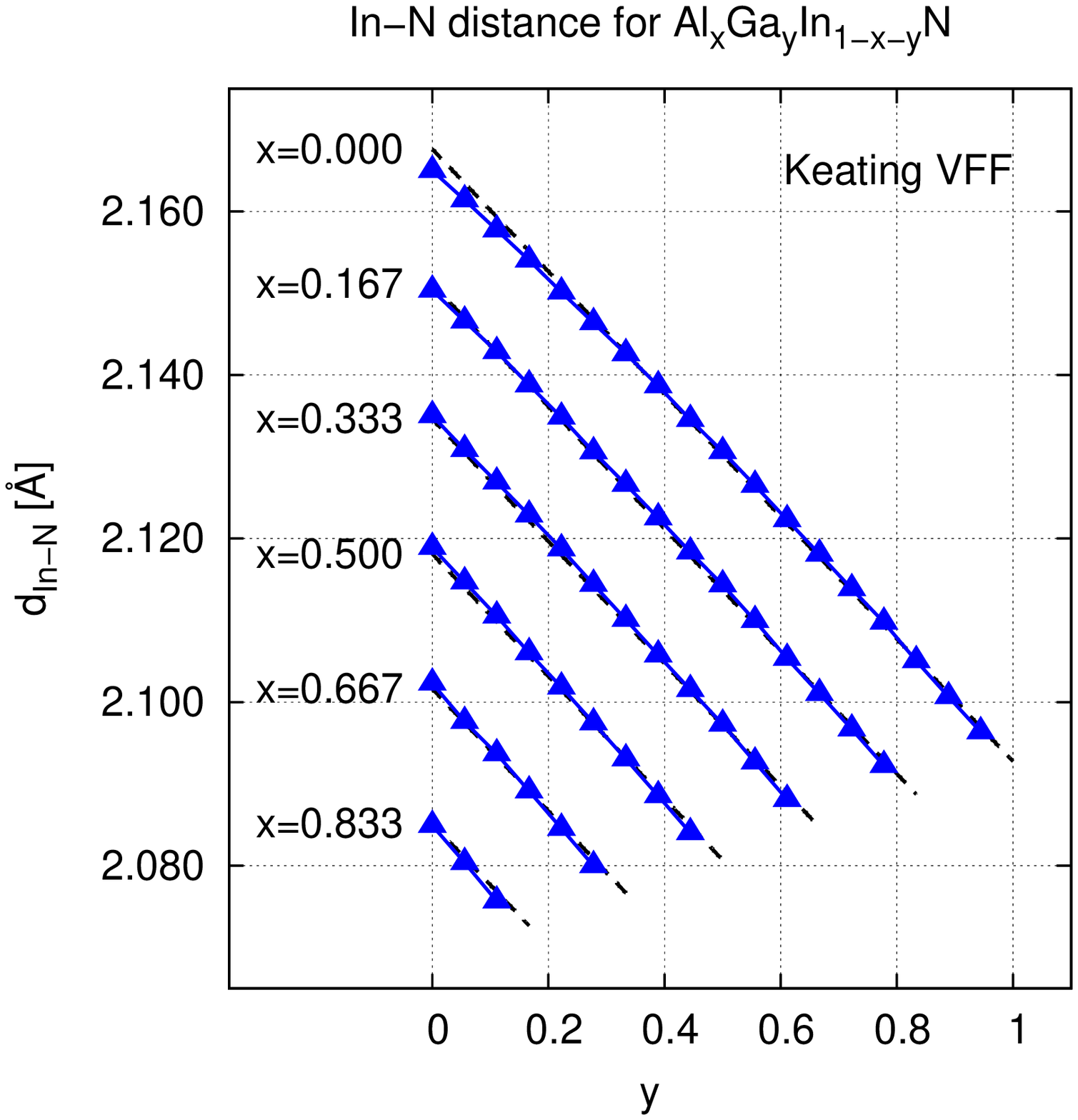}
	\includegraphics[height=0.5\textwidth,angle=270]
        {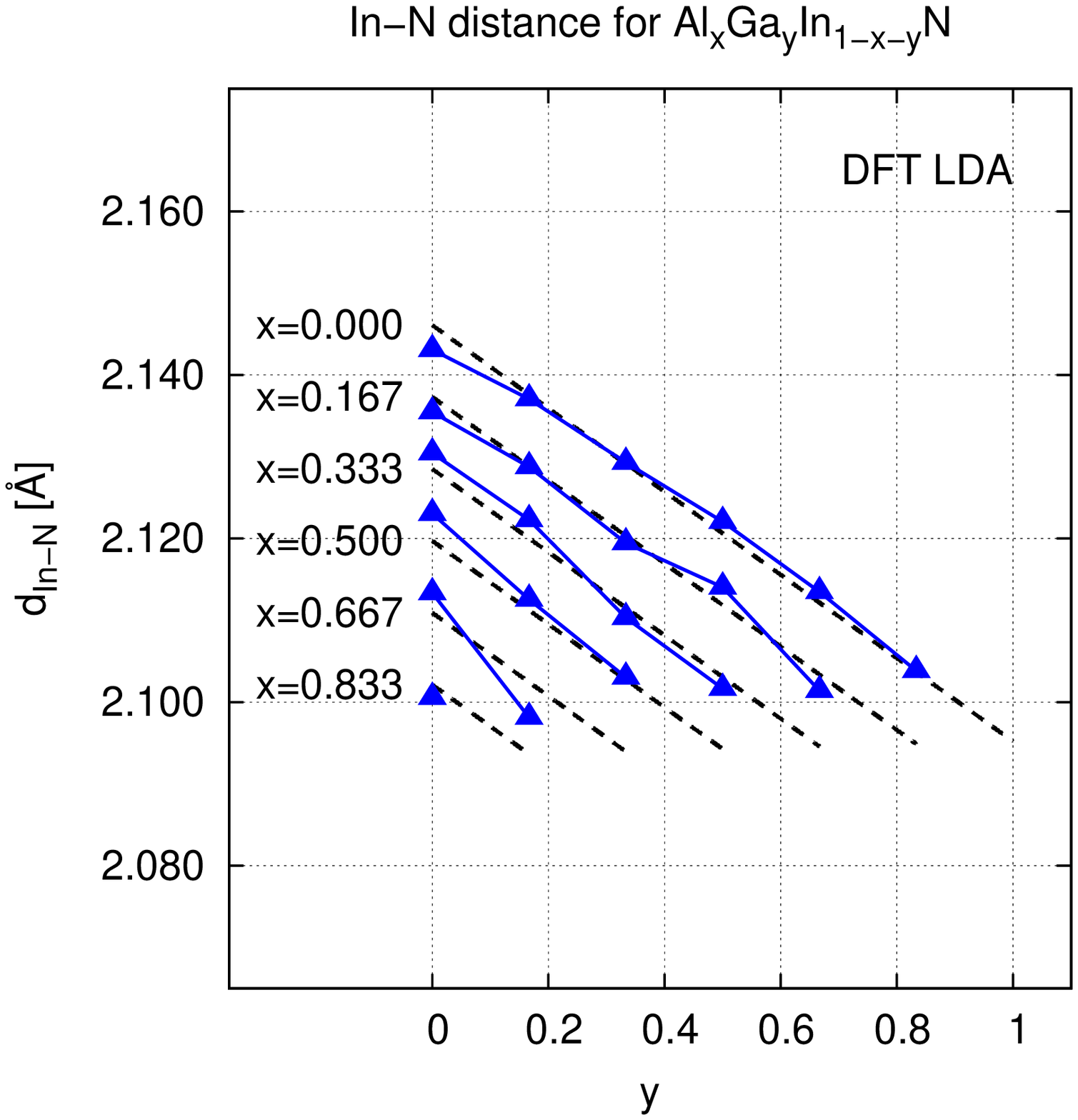}
	\caption{
	\label{Fig:BondLengths}
	The dependence of the average bond length on alloy composition for \AlGaInN
	obtained using Keating VFF and DFT LDA schemes. Points correspond to
	results of calculations, solid lines are only to guide the eye. Dashed
	lines denote linear fits presented in the table \ref{Tab:BondLengthsFits}.
	Note that for convenient comparison the scale on Keating VFF and DFT LDA
	graphs is the same.  }
\end{figure}

\begin{figure}[!ht]
	\includegraphics[height=0.50\textwidth,angle=270]
			{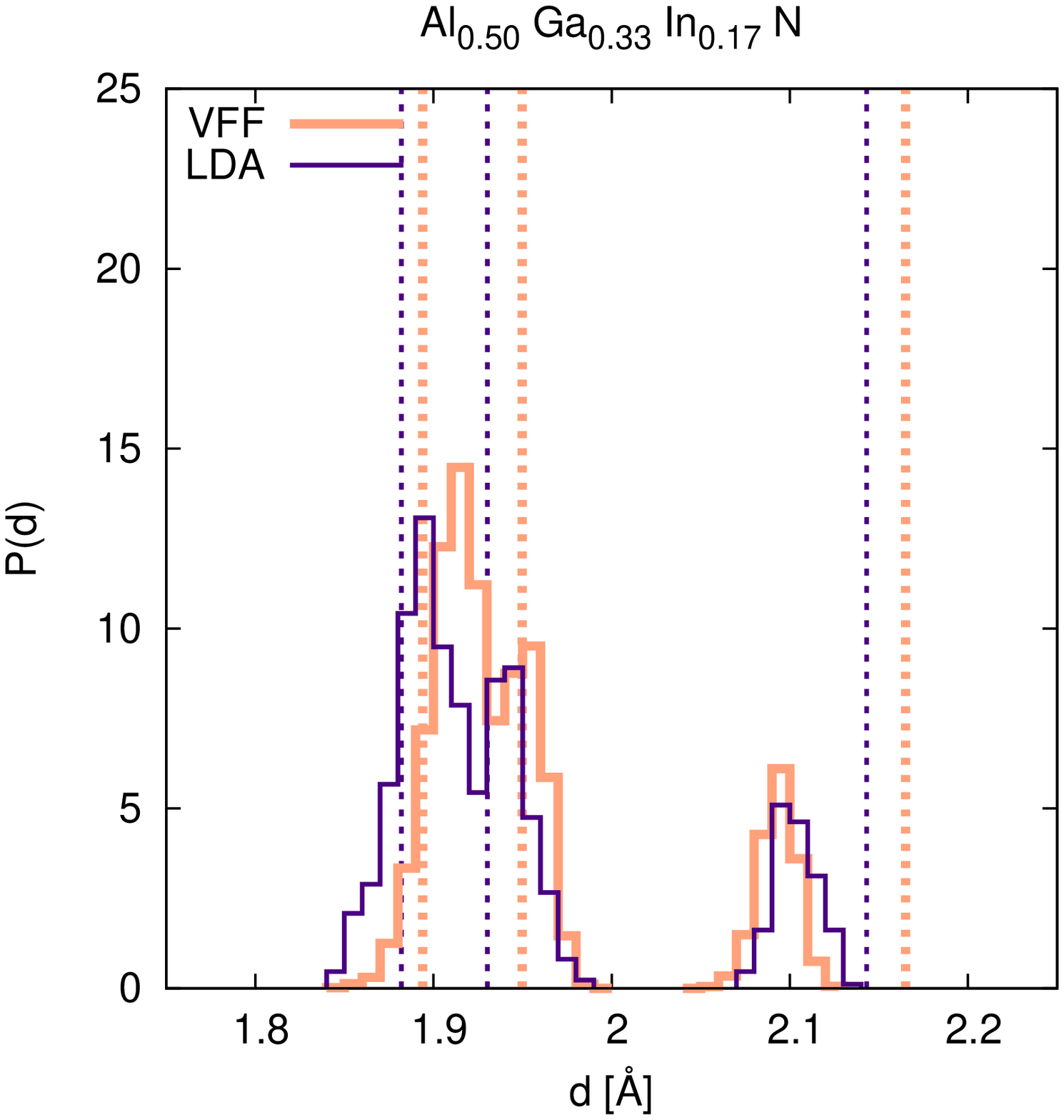} 
	\includegraphics[height=0.50\textwidth,angle=270]
			{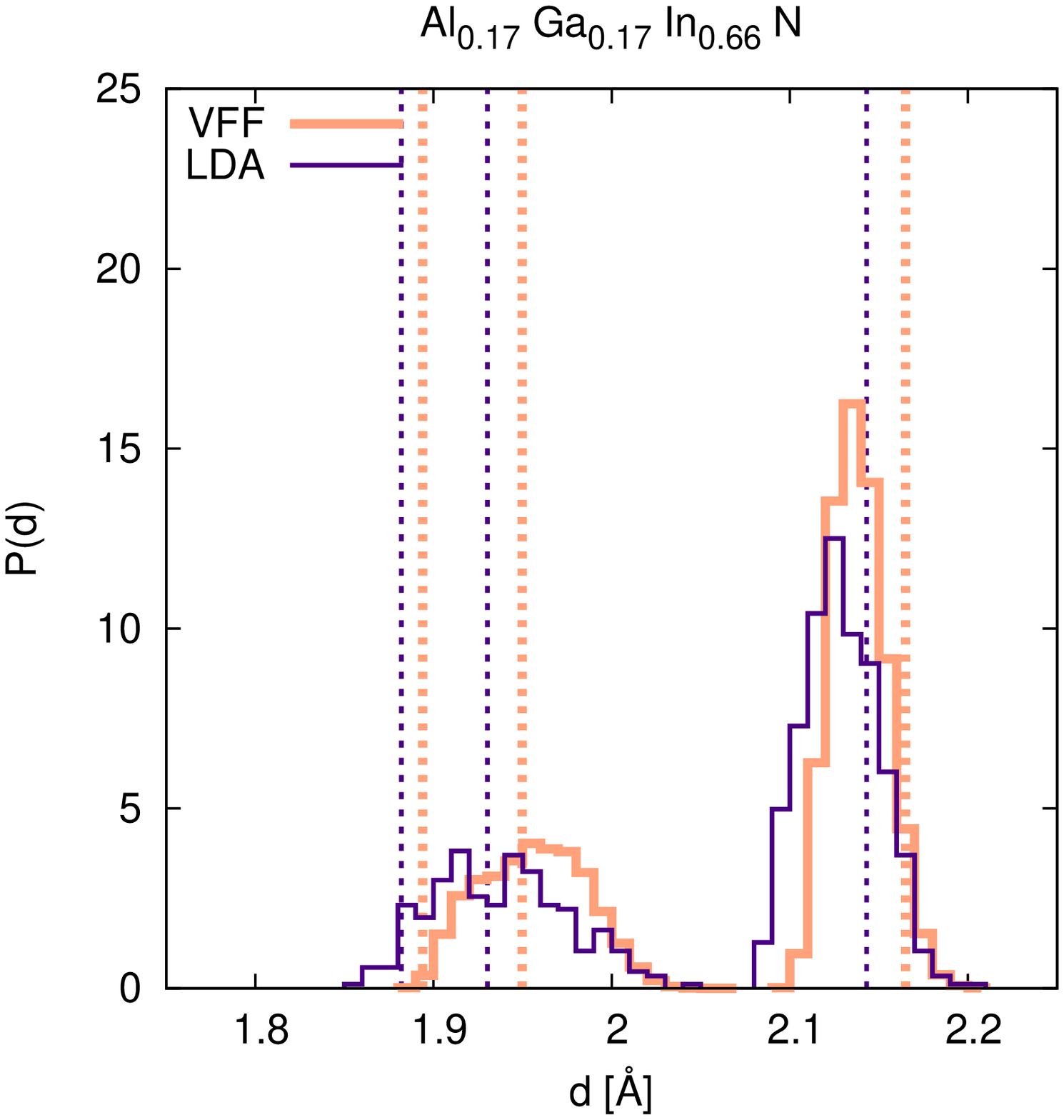} 
	\caption{\label{Fig:BondHistograms} 
	The nearest neighbour distance distribution resulting from Keating VFF and
	DFT LDA for Al$_{0.50}$Ga$_{0.33}$In$_{0.17}$N and
	Al$_{0.17}$Ga$_{0.17}$In$_{0.66}$N random alloys. Vertical dashed lines
	denote nearest neighbour distance in pure AlN, GaN and InN respectively as
	predicted by both presented models.  }
\end{figure}

\begin{figure}[!ht]
	\includegraphics[height=0.50\textwidth,angle=270]
			{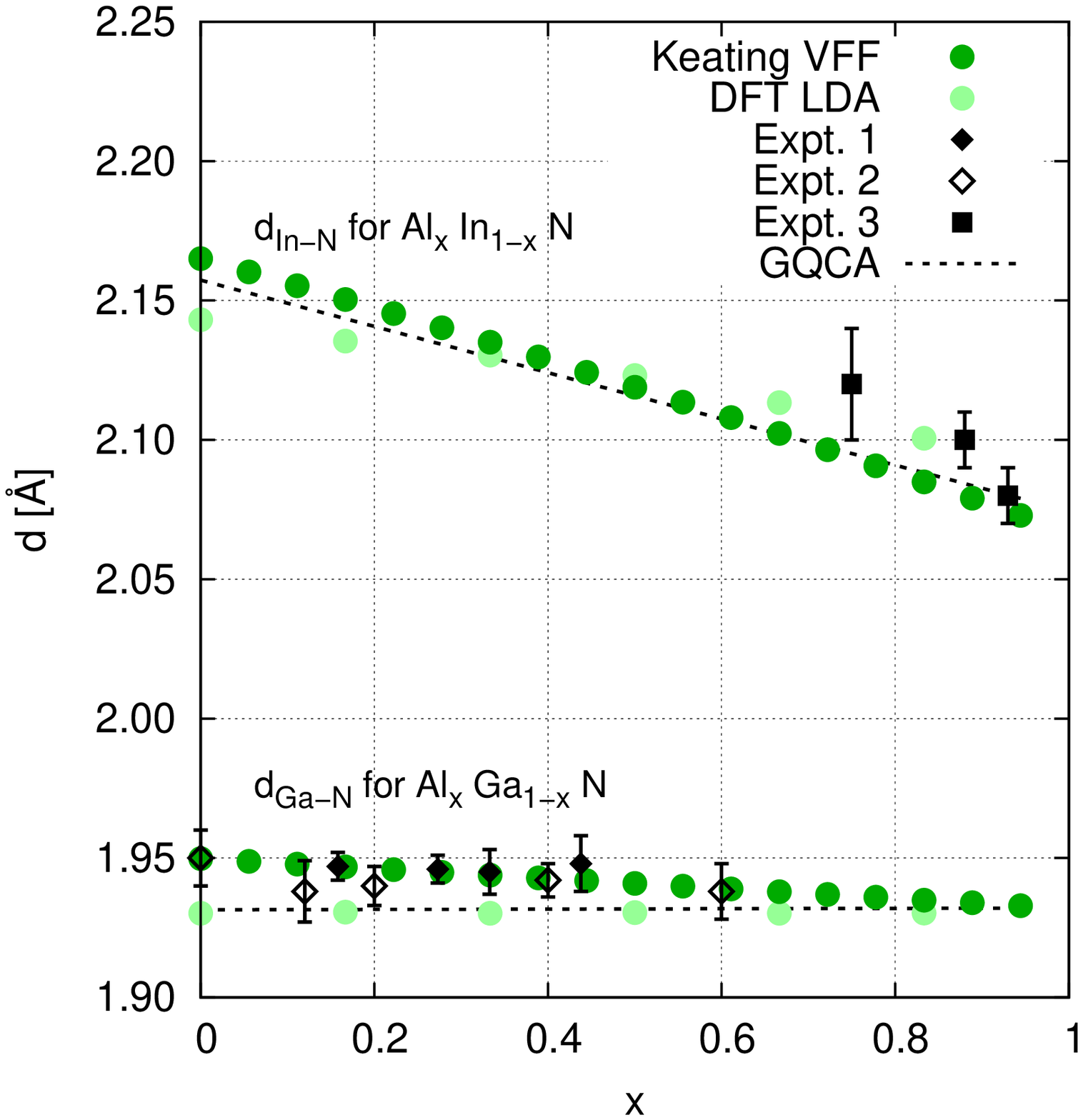} 
	\includegraphics[height=0.50\textwidth,angle=270]
			{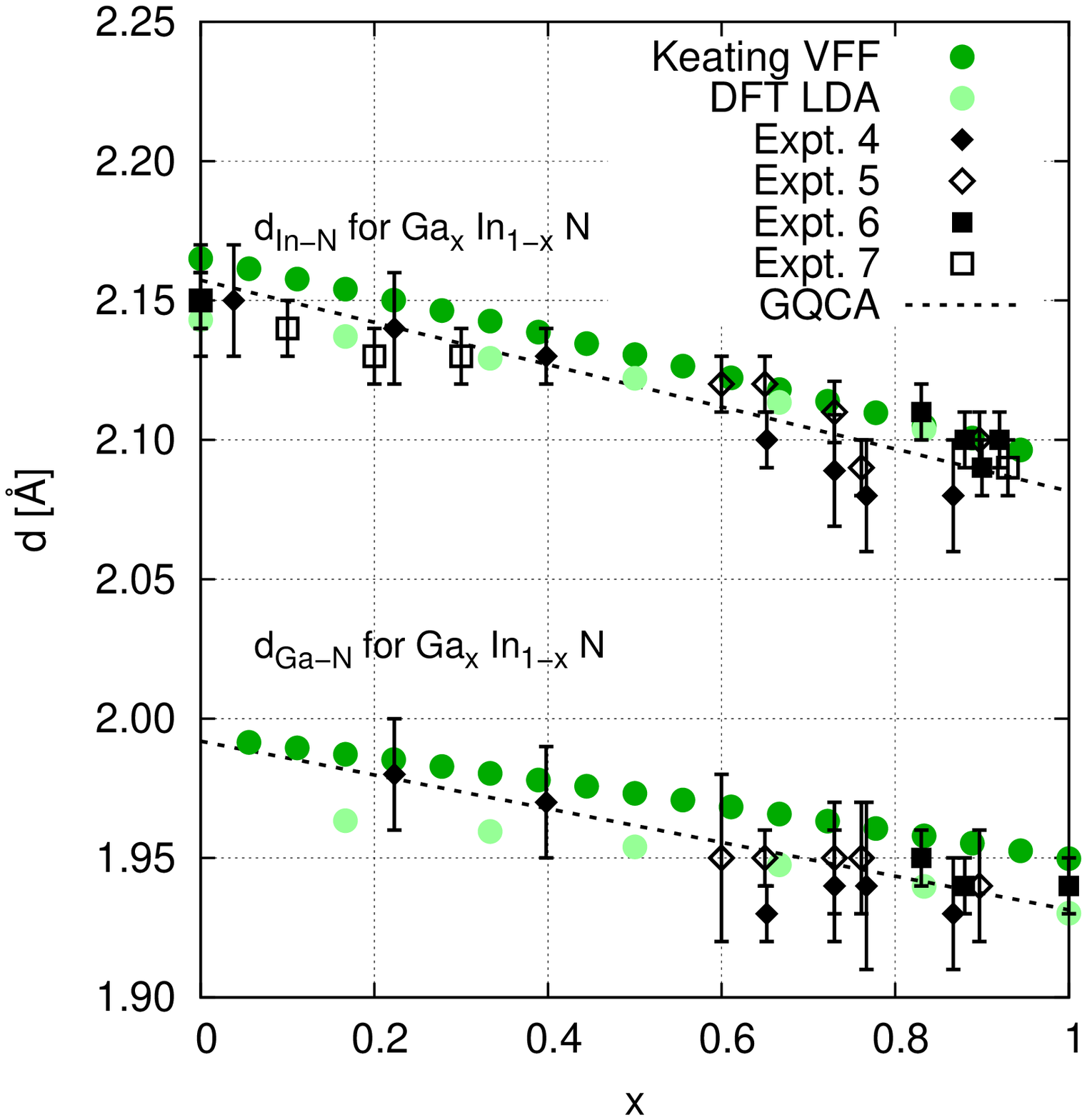}
\caption{\label{Fig:ComparisonWithExpt} Comparison of our theoretical results
obtained within both DFT LDA and Keating VFF models with various experimental
data. There are used data from the following experimental:
for \AlGaN 
Expt. 1 - Miyano \etal \cite{Miyano1997}, 
Expt. 2 - Yu \etal \cite{Yu1999}, 
for \AlInN 
Expt. 3 - Katsikini \etal \cite{Katsikini2008}, 
for \GaInN
Expt. 4 - MBE samples of Kachkanov \etal \cite{Kachkanov2006},
Expt. 5 - MOCVD samples of Kachkanov \etal \cite{Kachkanov2006}, 
Expt. 6 - Katsikini \etal \cite{Katsikini2003},
Expt. 7 - Katsikini \etal\cite{Katsikini2008}.
The results of previous theoretical predictions within GQCA \cite{Marques2006}
are presented for completeness. }
\end{figure}

Finally, it is interesting to examine, how the presented theoretical results
agree with experiments. We are not aware of any experimental results for bond
length distribution in \AlGaInN alloys, however, there is a significant body of
experimental data for ternaries \AlGaN, \AlInN, and \GaInN. Comparison of the
experimental bond lengths in ternary alloys with predictions of the Keating VFF
model is presented in figure \ref{Fig:ComparisonWithExpt}.  Even though
measurements are performed for wurtzite samples, the comparison with the
theoretical predictions for cubic phases is reasonable, owing to the local
similarity of both crystallographic phases.  In wurtzite structures, the
nearest neighbours bond elongations in the direction of crystallographic
$c$-axis are typically of the order of 0.01 $\AA$
\cite{Gorczyca2009a,Ambacher2002}. This is very similar to a typical
experimental error bars or bond length spread resulting from disorder in alloy
(compare histogram in figure \ref{Fig:BondHistograms}). As can be seen in
figure \ref{Fig:ComparisonWithExpt}, the comparison of available data with our
theoretical results reveals good agreement. It could be, of course, interesting
to see how this relates to the structural properties of cubic \AlGaInN which
growth using molecular beam epitaxy was recently reported \cite{As2007}. For
completeness, in figure  \ref{Fig:ComparisonWithExpt} we have also included the
DFT findings which again show the well known systematic tendency to
underestimate the bond lengths. 

\clearpage

\subsection{Next nearest neighbour distance}

Here we analyse the closest distances between the atoms of the same (cationic
or anionic) sublattice, i.e., the next nearest neighbour distance of an anion
or cation. The results obtained within both VFF and DFT models for a
nitrogen-nitrogen pairs are presented in figure \ref{Fig:DistNNNSummary}.
Qualitatively, the results for the smallest distances of cationic pairs are
very similar, exhibiting dominantly linear dependence on concentration of
constituents.  The coefficients $A$ and $B$ that determine this linear
dependence (see equation (\ref{Eq:LinearFit})) are presented in the table
\ref{Tab:DistNNNSummaryFits} for both cation-cation and nitrogen-nitrogen
average distance. The maximum error of the fit is also included there.  Since in
modern EXAFS experiments it is possible to measure the cation-cation distances
for every distinct pair separately, we have also performed the fits for all
possible combinations of cation-cation distances, namely Al-Al, Al-Ga, Al-In,
Ga-Ga, Ga-In, In-In (see table \ref{Tab:DistNNNCationsFits}).  Again, in all
cases linear model provides satisfactory description. 

\begin{table}[h!]
\caption{\label{Tab:DistNNNSummaryFits}
		 Results of linear fit to the average next nearest neighbour
		 cation-cation (CT - CT) and nitrogen-nitrogen (N - N) distances for
		 \AlGaInN quaternary alloys.}
\vspace{0.2cm}
\begin{indented}
\item[] \begin{tabular}{l|c|c|c}
\multicolumn{1}{c|}{Distance} & Model & Fit Function & Max. Error [\%] \\
\hline
CT-CT 
      & VFF     & $3.5147 - 0.43234\,x -0.33500\,y$ & 0.59\\ 
      & DFT LDA & $3.4934 - 0.42243\,x -0.34107\,y$ & 0.18 \\
\hline
N-N   
      & VFF     & $3.5200 - 0.43517\,x -0.33931\,y$ & 0.44 \\ 
      & DFT LDA & $3.4991 - 0.42487\,x -0.34562\,y$ & 0.07 \\
\end{tabular}
\end{indented}
\end{table}

\begin{table}[h!]
\caption{\label{Tab:DistNNNCationsFits}
         Results of linear fits to the average next nearest neighbour distances
         between various pairs of cations for \AlGaInN quaternary alloys.}
\vspace{0.2cm}
\begin{indented}
\item[] \begin{tabular}{l|c|c|c}
\multicolumn{1}{c|}{Distance} & Model & Fit Function & Max. Error [\%] \\ 
\hline
Al-Al & VFF     & $3.3660 - 0.27416\,x - 0.20666\,y$ & 0.15 \\ 
      & DFT LDA & $3.3663 - 0.29078\,x - 0.23771\,y$ & 0.78 \\
\hline
Al-Ga & VFF     & $3.3865 - 0.28385\,x - 0.21459\,y$ & 0.07 \\ 
      & DFT LDA & $3.3798 - 0.29593\,x - 0.23651\,y$ & 0.24 \\ 
\hline
Al-In & VFF     & $3.4599 - 0.32354\,x - 0.24685\,y$ & 0.05 \\
      & DFT LDA & $3.4286 - 0.30625\,x - 0.23513\,y$ & 0.31 \\ 
\hline
Ga-Ga & VFF     & $3.4057 - 0.29338\,x - 0.22179\,y$ & 0.18 \\ 
      & DFT LDA & $3.3922 - 0.29127\,x - 0.23931\,y$ & 0.43 \\ 
\hline
Ga-In & VFF     & $3.4771 - 0.33479\,x - 0.25554\,y$ & 0.06 \\
      & DFT LDA & $3.4474 - 0.30469\,x - 0.24991\,y$ & 0.26 \\ 
\hline
In-In & VFF     & $3.5362 - 0.38725\,x - 0.29734\,y$ & 0.11 \\ 
      & DFT LDA & $3.4983 - 0.31251\,x - 0.26031\,y$ & 0.53 \\ 
\end{tabular}
\end{indented}
\end{table}

In addition to the dependence of average cation-cation and anion-anion distance
on concentration, we have also analyzed the shape of distribution. Sample
histograms are depicted in figure \ref{Fig:DistNNNHistogram}.  When inspecting
presented graphs, one can notice that cationic and nitrogen sublattices relax
in a different manner.  The behaviour of cation-cation distribution is similar
to virtual crystal exhibiting unimodal shape.  At the same time
nitrogen-nitrogen distance is much more distorted with respect to single-peaked
virtual crystal picture. The shape of the distribution is multimodal. The peaks
correspond to three possible combinations of N-N pair joined by Al, Ga or In
atom respectively. Since the equilibrium distances of Al-N and Ga-N are very
similar the pairs of N-N joined by Al and Ga form common maximum on the
presented plot.  These findings agree very well with the same behaviour
reported for ternaries \cite{Schabel1991,Mattila1999}.
Comparing bond length histograms obtained within the VFF and DFT schemes (see
figure \ref{Fig:DistNNNHistogram}), it is interesting to note that the
agreement in the predicted next nearest neighbour distances is in this case
even better than for the nearest neighbours histograms. To some extent it is
caused by the fact that the larger peak widths than in the case of the nearest
neighbour distance distributions cause that systematic differences between bond
length predicted by DFT and VFF are less pronounced. Since the VFF model
includes explicit interactions only between the nearest neighbours, one could
think that the description of second coordination shell would be less accurate.
However, it turns out that the VFF model provides also very reliable
predictions for cation-cation and anion-anion distances (see figure
\ref{Fig:DistNNNHistogram}).

Even though the experimental data for quaternaries are unavailable, similarly
like in the nearest neighbour case, there is a considerable number of results
for ternaries. The comparison of experimental findings with our theoretical
results are presented in figure \ref{Fig:ComparisonWithExptNNN}. Generally, the
agreement is very good, however, a few things are worth pointing out here. One
can notice that quite often the  results measured by different groups exhibit
significant spread and, in addition, some of the experimental data exhibit
large error bars.  This underlines the fact that such measurements are on the
verge of available experimental technique.  The largest discrepancies between
theory and experiment we observe for \AlInN alloy (see the middle column of the
graph \ref{Fig:ComparisonWithExptNNN}), however, there is only one recent
report \cite{Katsikini2008} known to us, which deals with the structure of this
material. To elucidate the matter, more experimental data would be very
helpful.   One has also to bear in mind that our calculation assume random
cation distribution in the sample. If some kind of clustering for particular
type of cations occurs, it could lead to modification of the presented results. 

\begin{figure}[!ht]
 	\includegraphics[height=0.45\textwidth,angle=270]
			{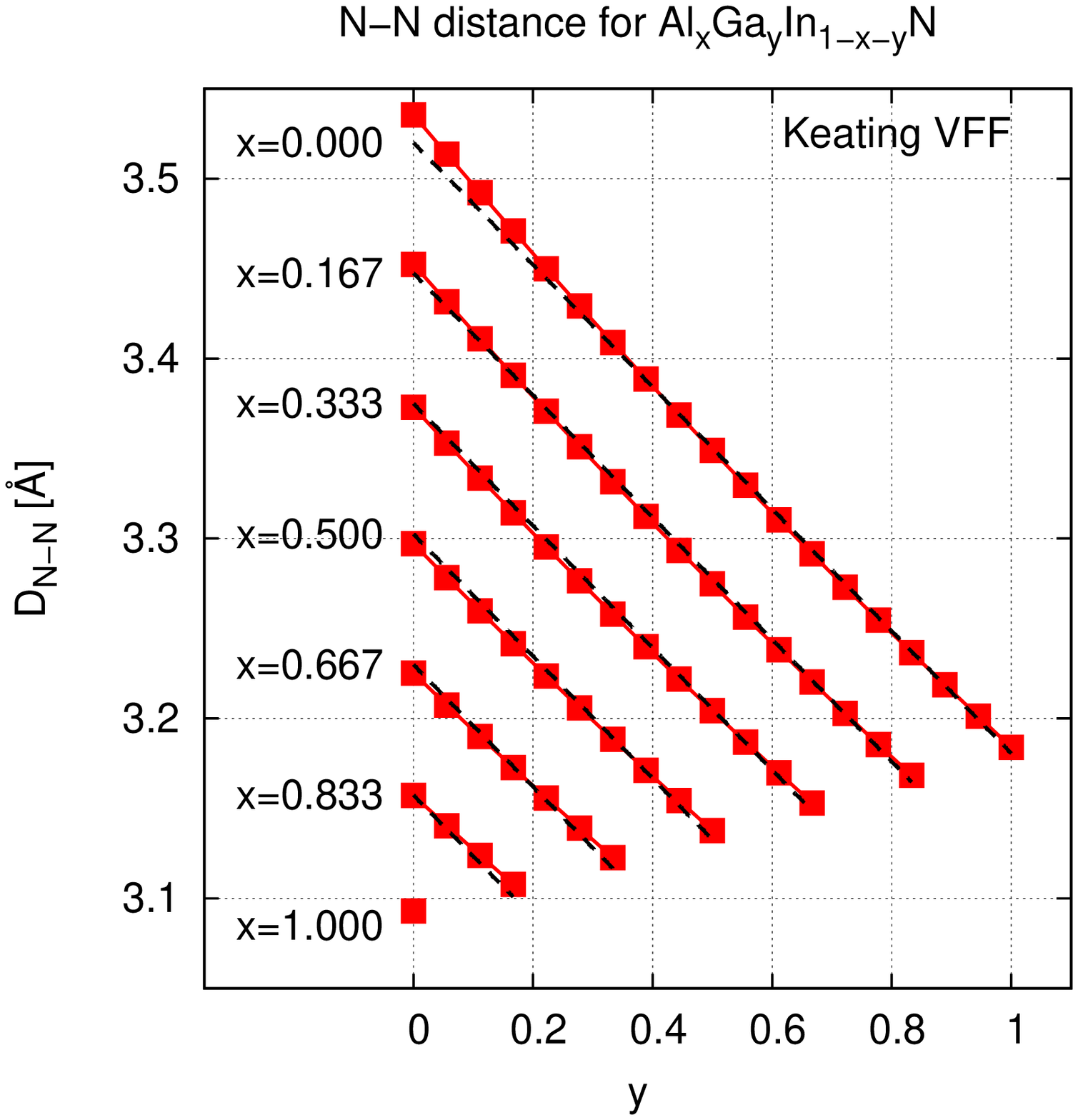}
	\includegraphics[height=0.45\textwidth,angle=270]
			{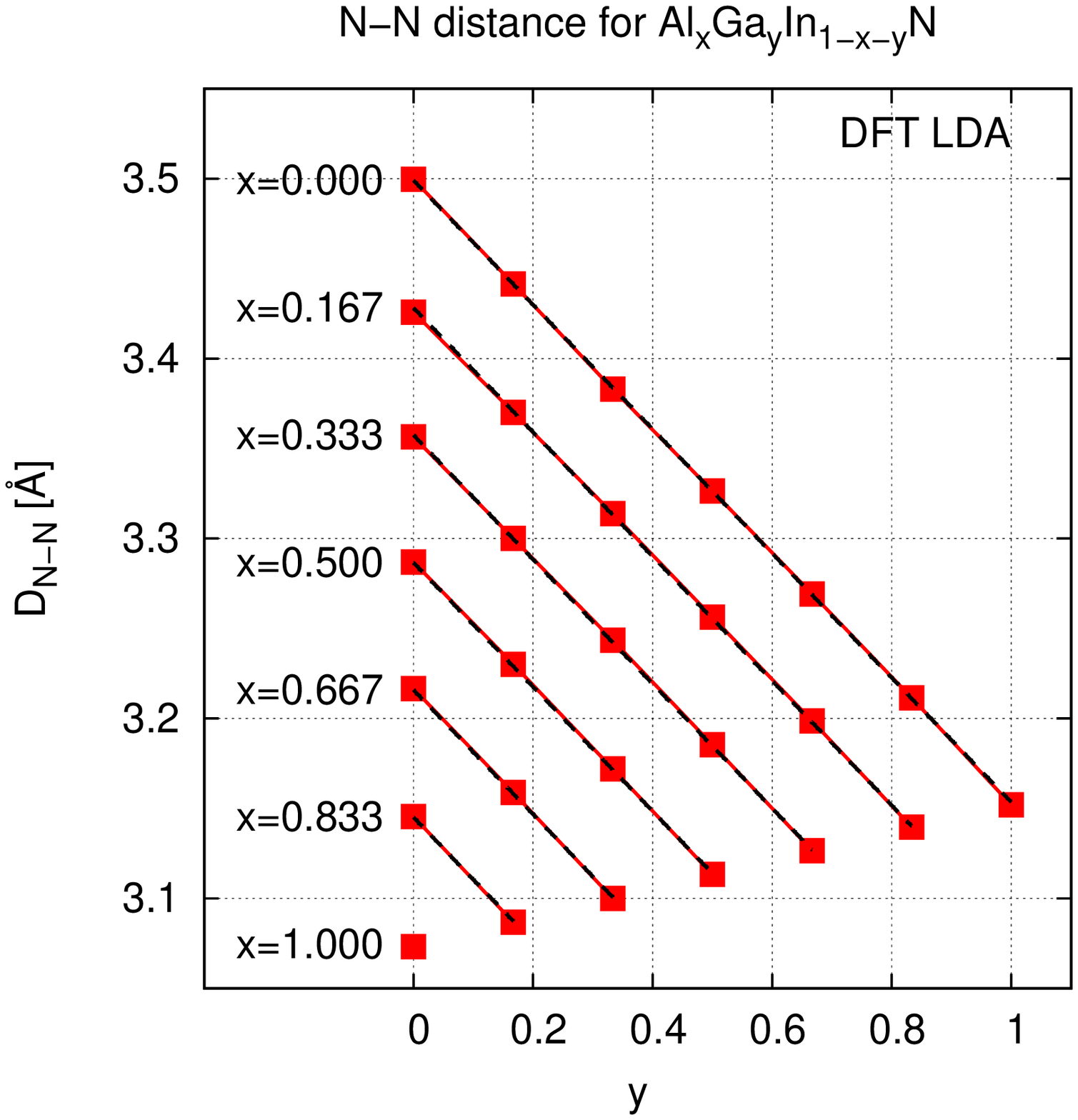}
			
	\caption{
 	 \label{Fig:DistNNNSummary}
		 The nitrogen-nitrogen average next nearest neighbour distances as a
		 function  of composition for \AlGaInN obtained using Keating VFF and
		 DFT LDA, for the case of nitrogen-nitrogen.  Points correspond to
		 results of calculations, solid lines are only to guide the eye. Dashed
		 lines denote linear fits presented in the table
		 \ref{Tab:DistNNNSummaryFits}.  Note that for convenient comparison the
		 scale on all graphs is the same.
	}
\end{figure}

\begin{figure}[!ht]
	\includegraphics[height=0.45\textwidth,angle=270]
			{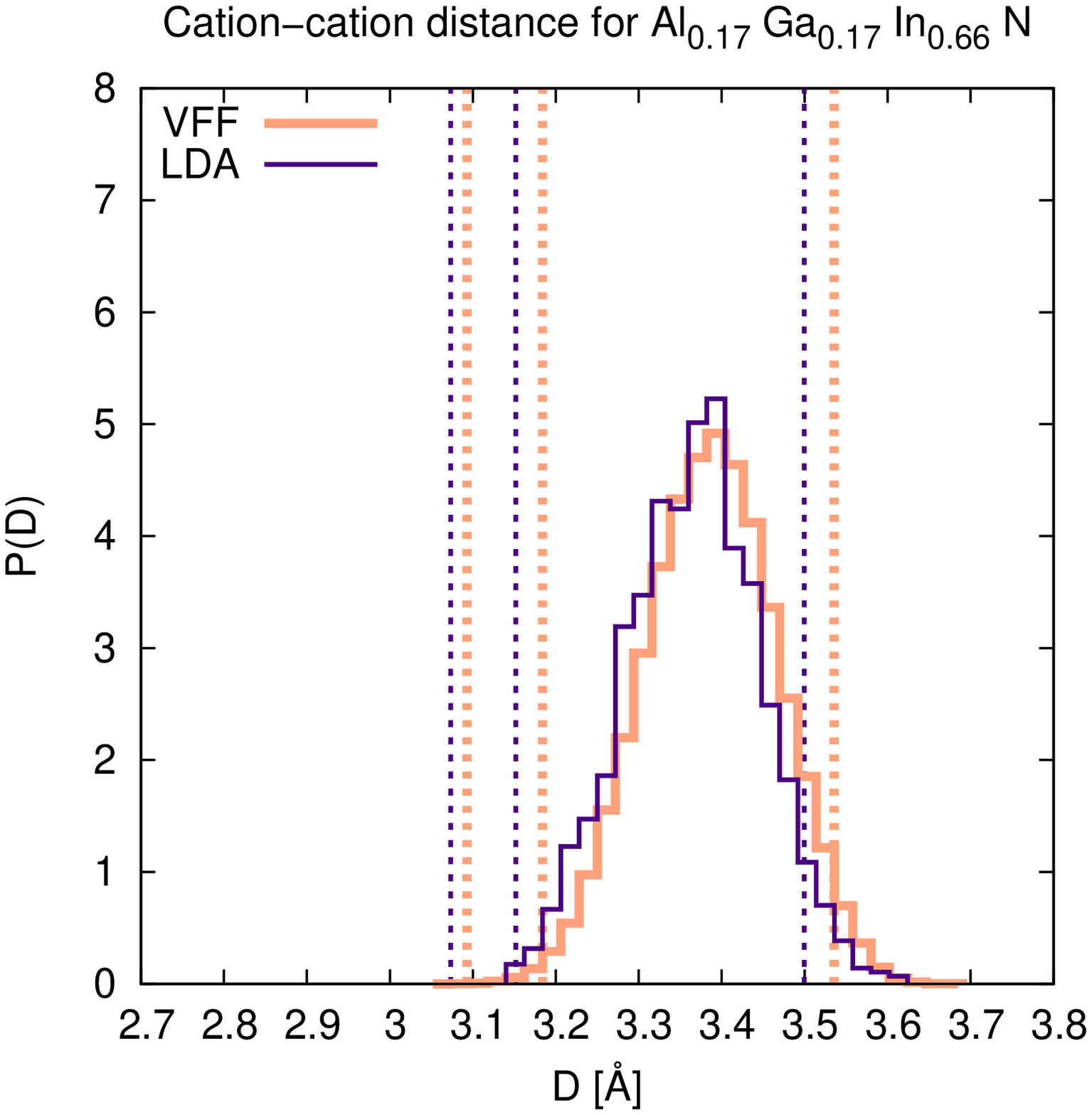}
	\includegraphics[height=0.45\textwidth,angle=270]
			{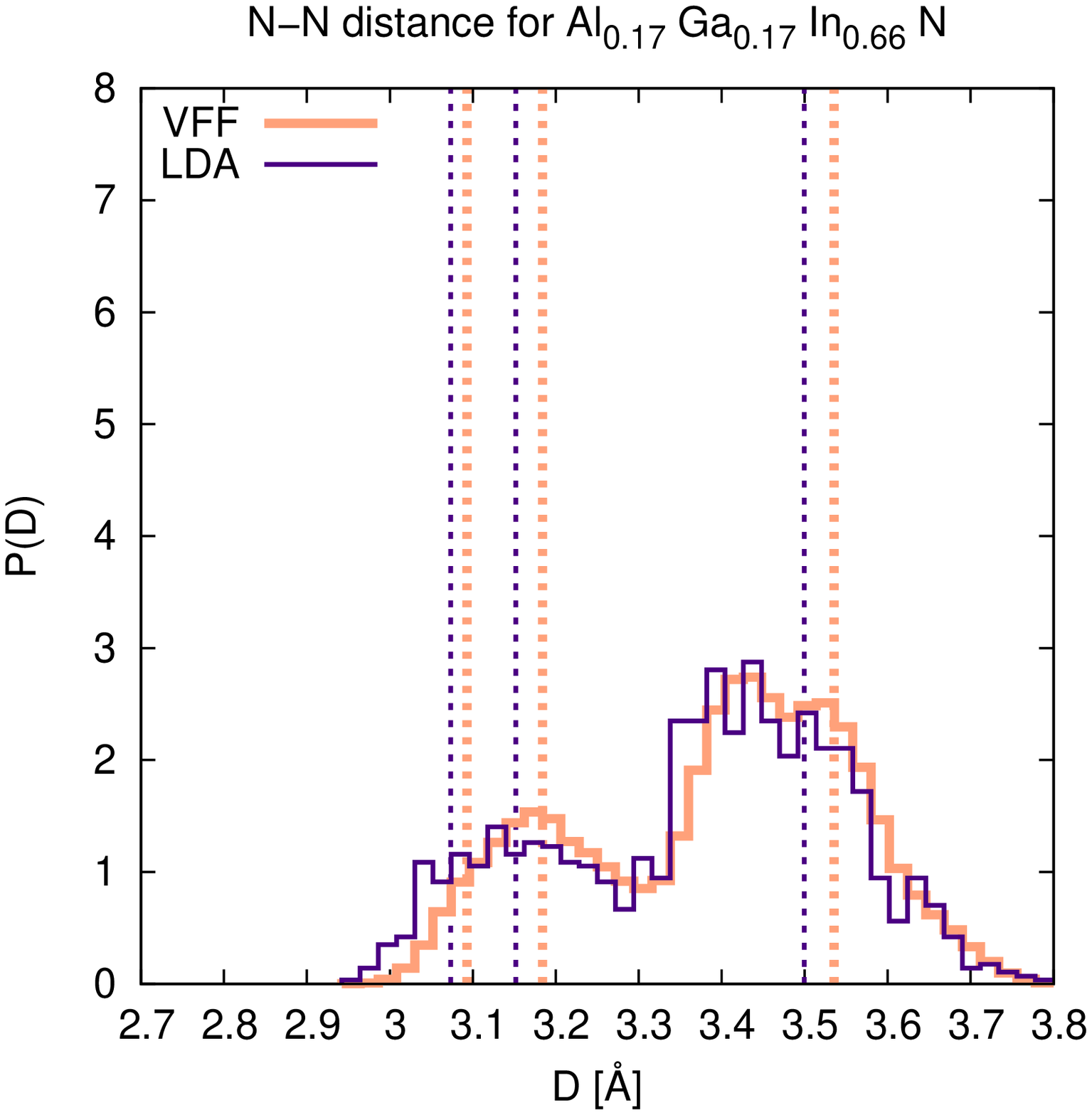}
\caption{
 	 \label{Fig:DistNNNHistogram}
	   Comparison of the next nearest neighbour distance distribution 
	   resulting from the Keating VFF and DFT LDA for 
	   Al$_{0.17}$Ga$_{0.17}$In$_{0.66}$N.
	   Vertical dashed lines denote next nearest neighbour distance in 
	   pure AlN, GaN and InN respectively as predicted by both presented 
	   models.
	}
\end{figure}

\begin{figure}[p]
	\includegraphics[height=0.33\textwidth,angle=270]
        {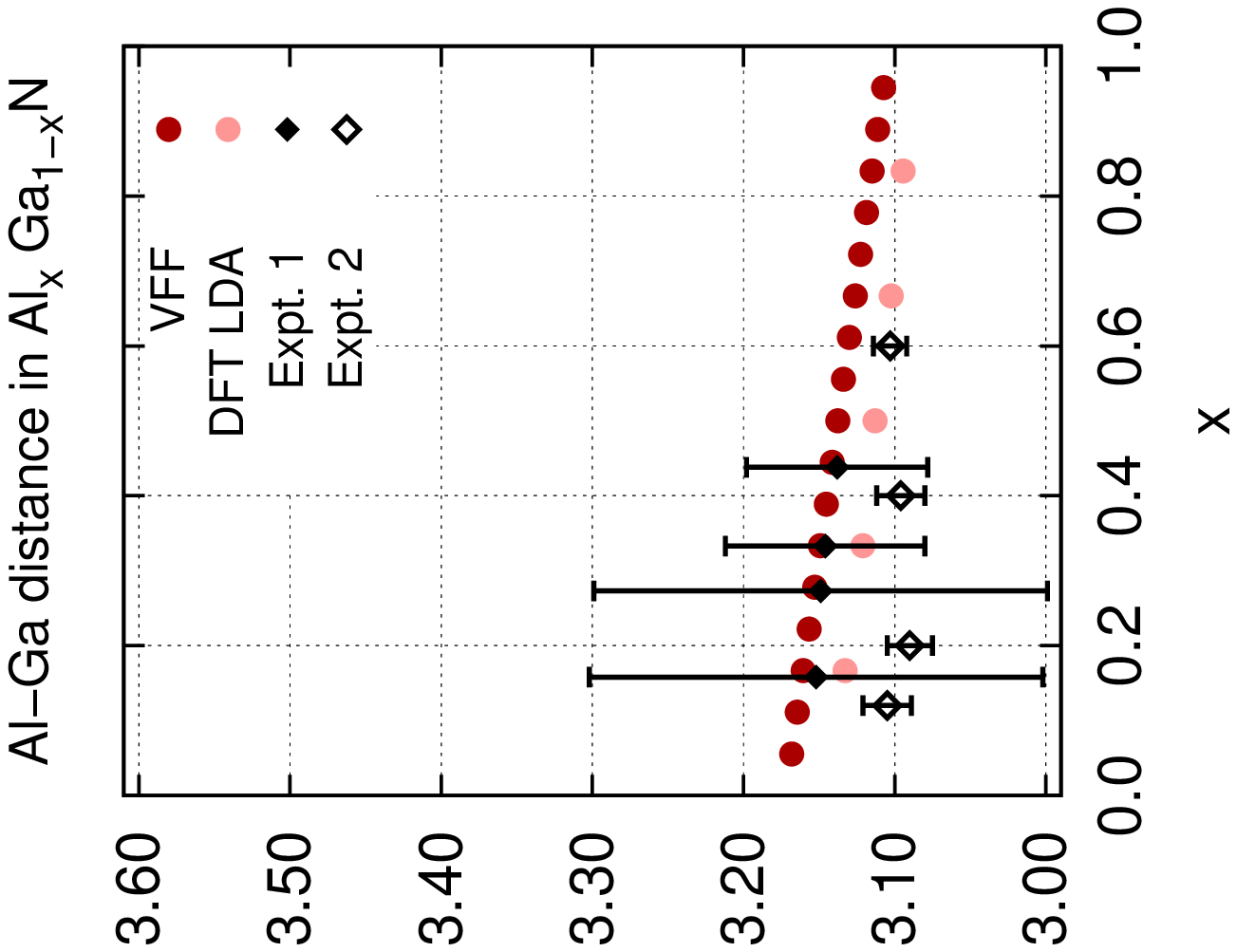}
        \includegraphics[height=0.33\textwidth,angle=270]
        {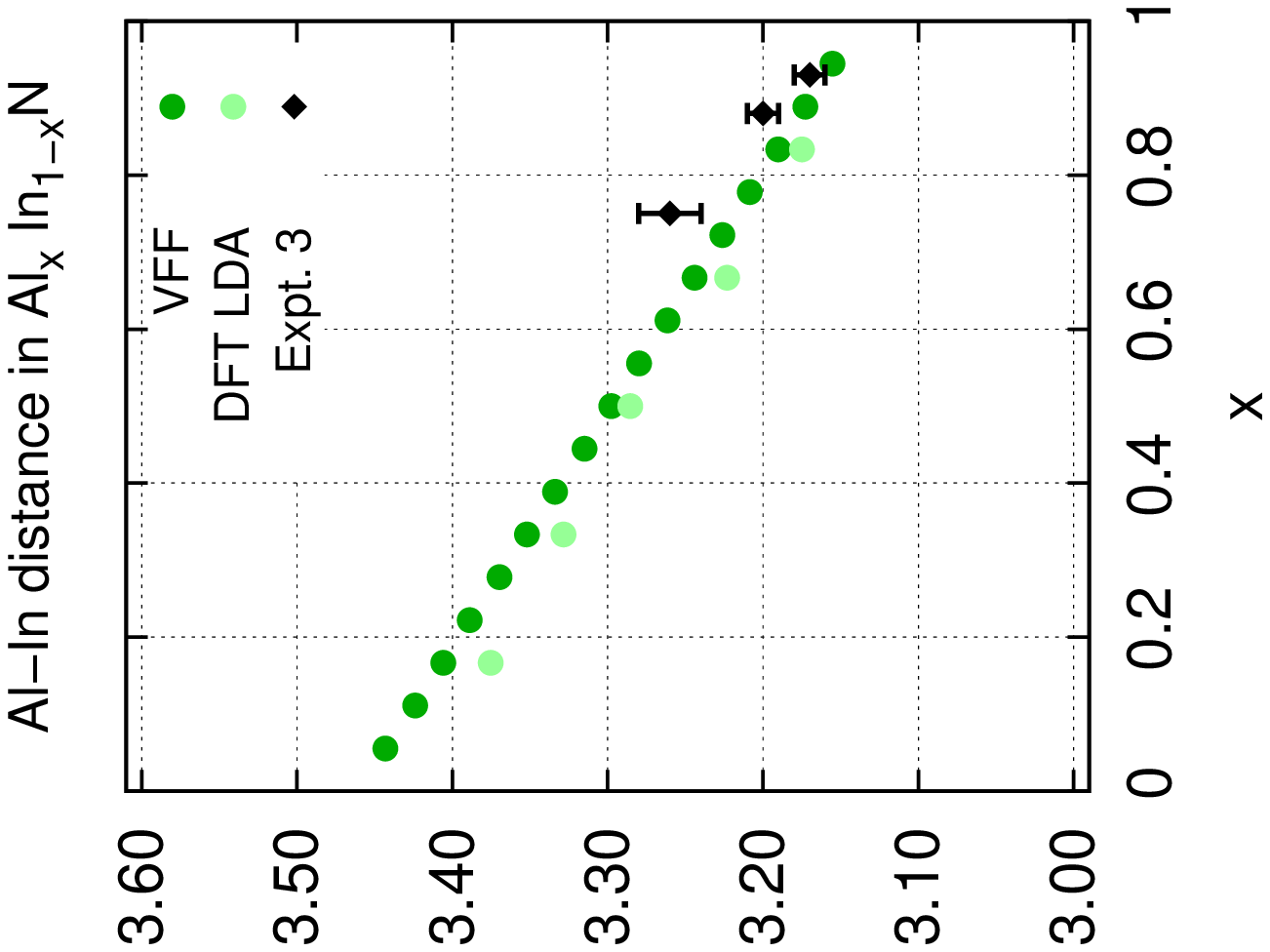}
	\includegraphics[height=0.33\textwidth,angle=270]
        {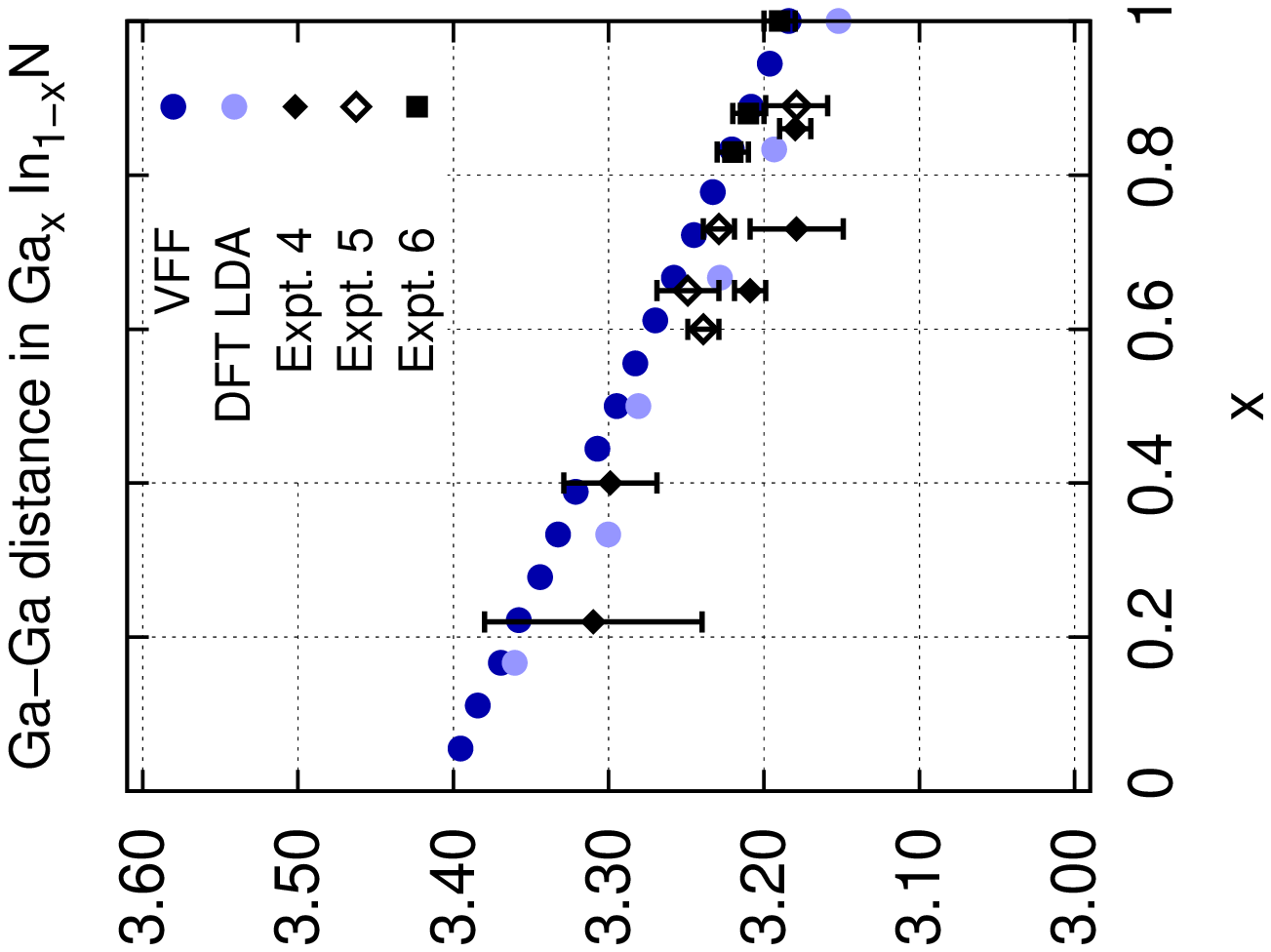}
	\includegraphics[height=0.33\textwidth,angle=270]
        {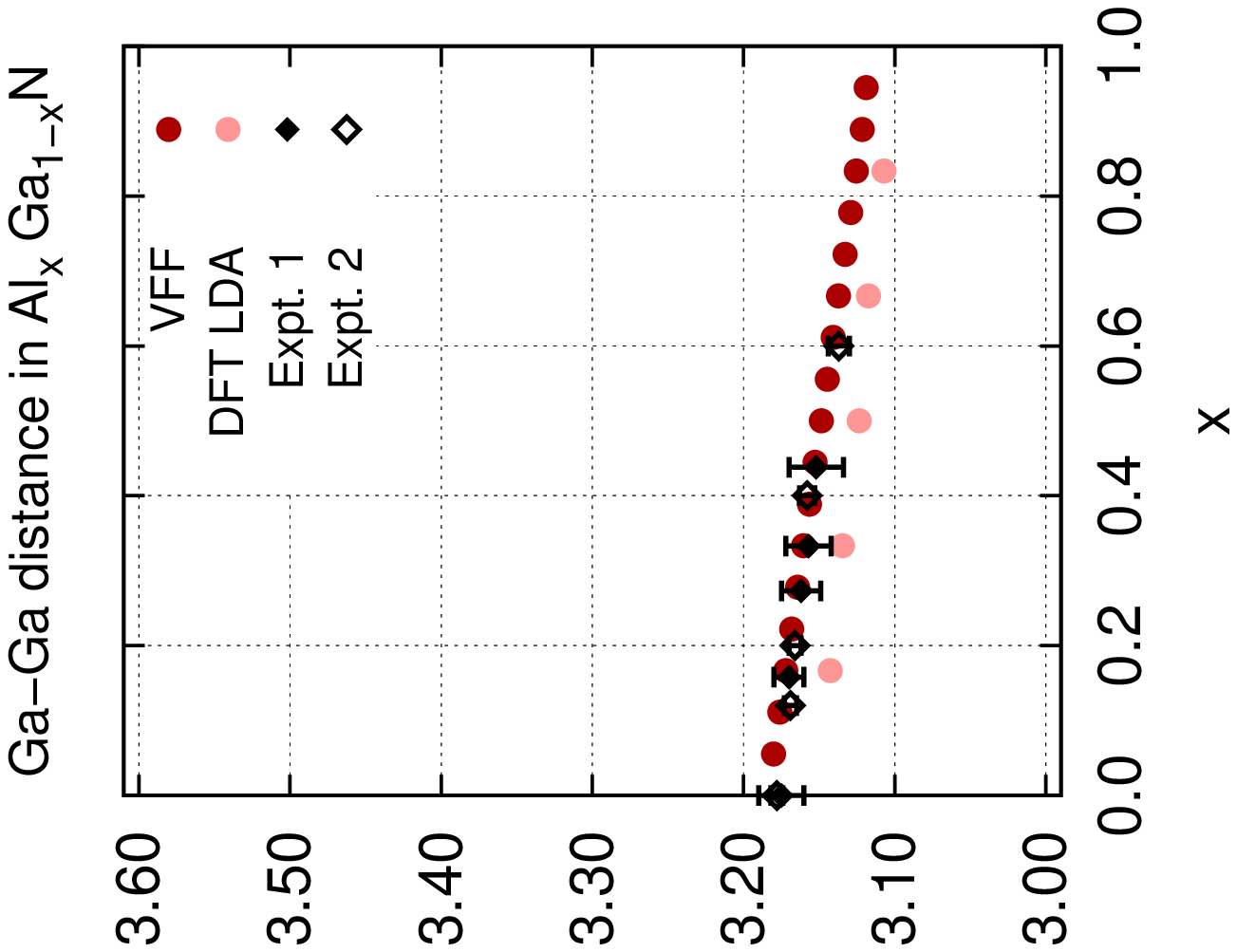}
        \includegraphics[height=0.33\textwidth,angle=270]
        {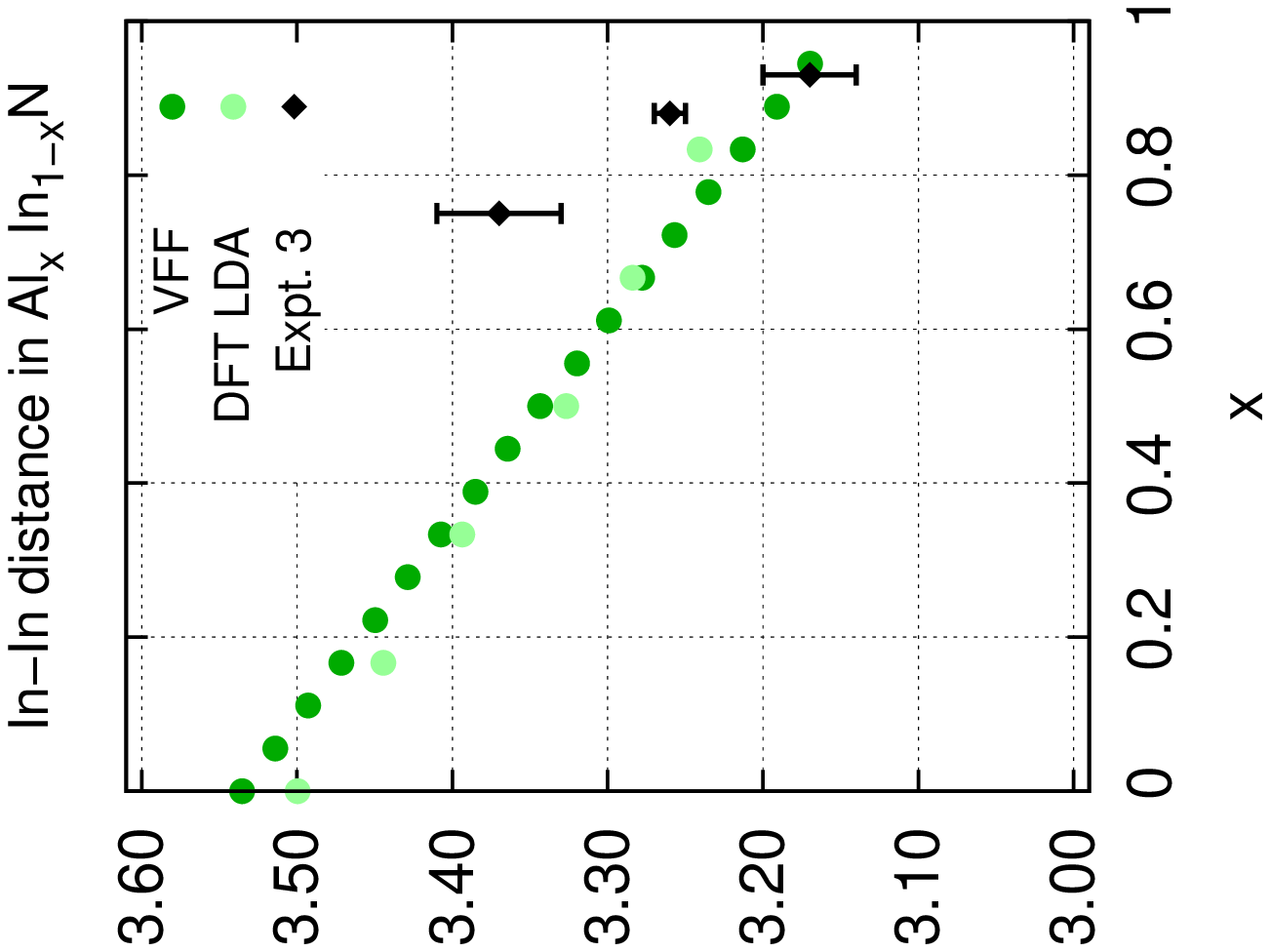}
	\includegraphics[height=0.33\textwidth,angle=270]
        {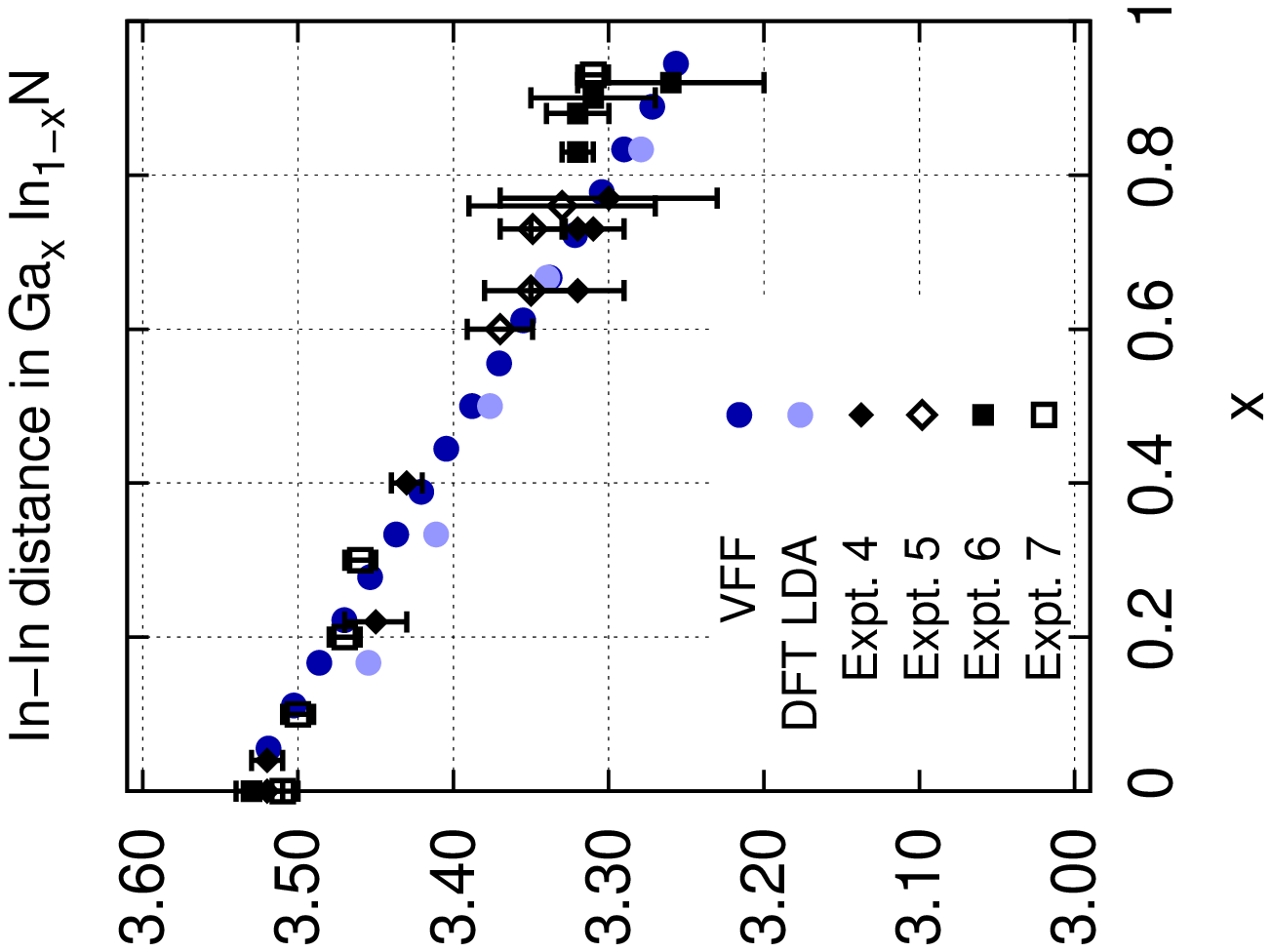} 
	\caption{
	\label{Fig:ComparisonWithExptNNN}
	Comparison of our theoretical results for average next nearest neighbours
	distances with various experimental findings:
	for \AlGaN
	Expt. 1 - Miyano \etal \cite{Miyano1997},
	Expt. 2 - Yu \etal \cite{Yu1999},
	for \AlInN
	Expt. 3 - Katsikini \etal \cite{Katsikini2008},
	for \GaInN
	Expt. 4 - MBE samples of Kachkanov \etal \cite{Kachkanov2006},
	Expt. 5 - MOCVD samples of Kachkanov \etal \cite{Kachkanov2006},
	Expt. 6 - Katsikini \etal \cite{Katsikini2003},
	Expt. 7 - Katsikini \etal\cite{Katsikini2008}.
	All distances are in Angstroms.
        } 
\end{figure}

\clearpage
 
\section{Elastic constants for alloys calculated using Keating model
\label{Sec:ElasticConstants}}

In this section we present results of calculations of  elastic constants
$c_{ij}$ for random \AlGaInN alloys.  Calculations were carried out using
Keating VFF and cover the whole concentration range. To extract the values of
the alloy elastic constants, we have applied three types of strains to every
alloy supercell:
\begin{eqnarray}
	\bm{\epsilon}_A &=& [ \epsilon, 0, 0, 0, 0, 0 ],    \nonumber \\
	\bm{\epsilon}_B &=& [ \epsilon, -\epsilon, 0, 0, 0, 0 ], \\
	\bm{\epsilon}_C &=& [  0, 0, 0, 0, 0, \epsilon ]. \nonumber
\end{eqnarray}
For each type of the deformation, $\epsilon$ was varied within the range of
values  $\{ - 1.0 \%, -0.5 \%, 0.5 \%, 1\% \}$ and the elastic energy has been
calculated. Then, on the basis of strain energy relation
\begin{equation}
 E = \frac{1}{V_0} \sum_{i,j=1}^6 c_{ij} \epsilon_i \epsilon_j,
\end{equation}
three elastic constants $c_{11}$, $c_{12}$ and $c_{44}$ have been determined
from parabolic fits to the energy for deformations $\bm{\epsilon}_A$,
$\bm{\epsilon}_B$, $\bm{\epsilon}_C$ .

The results are presented in figure \ref{Fig:ElasticConstants}.  We also
included there the prediction of Vegard-like law for elastic constants:
\begin{equation}
    c_{ij}^{\rm{Vegard}}(x,y)=x\, c_{ij}^{\rm{AlN}} + 
	                     y\, c_{ij}^{\rm{GaN}} +
						 (1-x-y) \, c_{ij}^{\rm{InN}}.
\end{equation}
Exact functional forms of this equation for $c_{11}$, $c_{12}$ and $c_{44}$
are explicitly given in the table \ref{Tab:ElasticConstantsAlloys}. However,
after a brief analysis of graph \ref{Fig:ElasticConstants}, one notices that
Keating VFF results are not very well described by the above Vegard's law. This
is particularly pronounced for elastic constants $c_{11}$ and $c_{44}$.  The
deviation for $c_{12}$ is very weak, but this elastic constant has also the
lowest discrepancy between materials AlN, GaN and InN.  To fully describe the
dependence of $c_{ij}$ on composition one has to include bowing term $\Delta
c_{ij}(x,y)$, which is defined as follows:
\begin{eqnarray}
	c_{ij}&=&c_{ij}^{\rm{Vegard}}(x,y) + \Delta\, c_{ij}(x,y), \\ 
	\Delta\, c_{ij}(x,y) &=& P\,x(1-x) + Q\, x y + R\,y(1-y). \nonumber
\end{eqnarray}
After including this additional function and performing fitting procedure for
$P$, $Q$ and $R$, the VFF results are reproduced with accuracy much better than
1 GPa. The coefficients $P$, $Q$, and $R$ for cubic elastic constants are
provided in the table \ref{Tab:ElasticConstantsAlloys}.

\begin{figure}[!ht]
	\begin{center}
	\includegraphics[height=0.49\textwidth,angle=270]
			{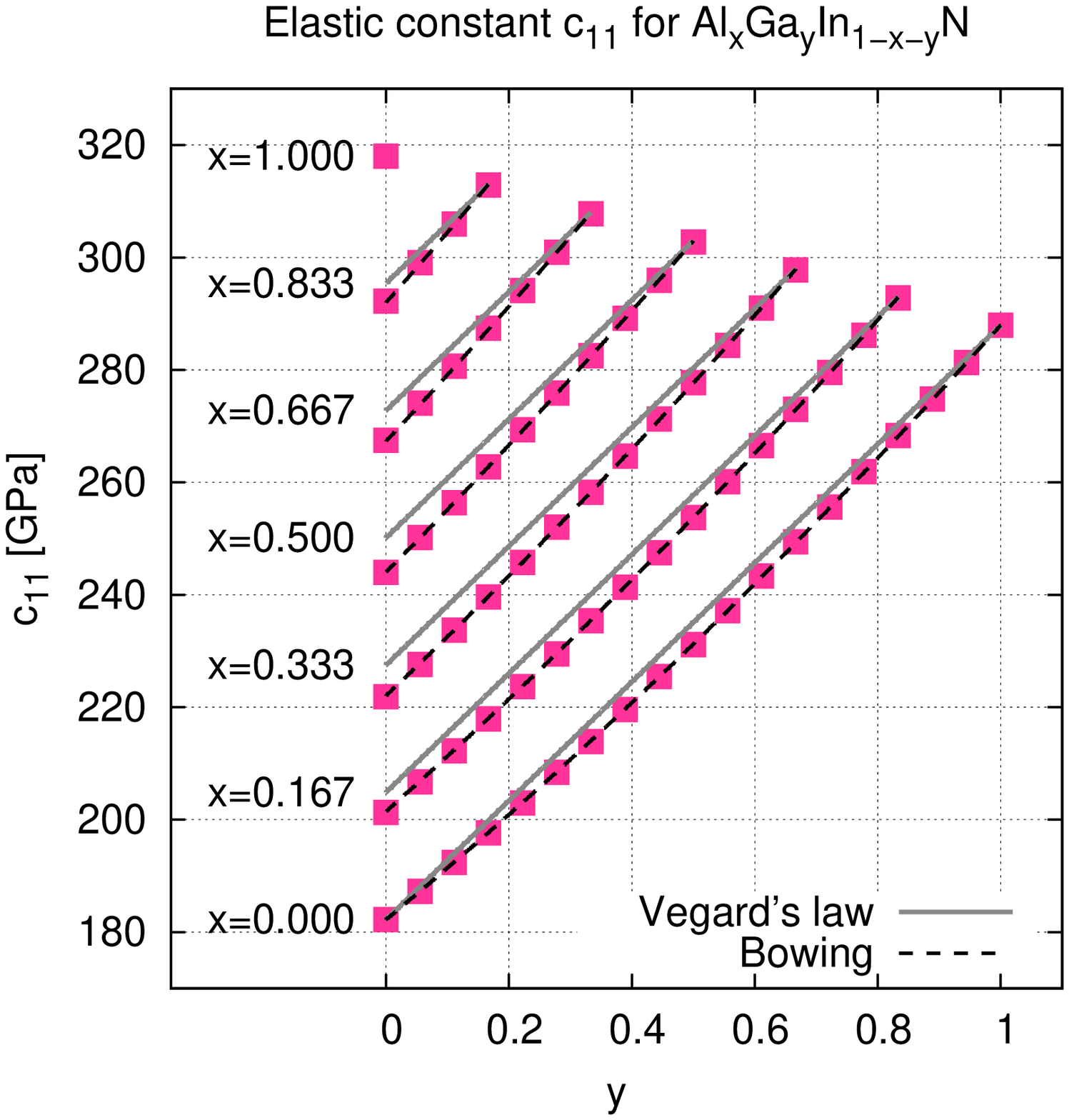} 
	\includegraphics[height=0.49\textwidth,angle=270]
			{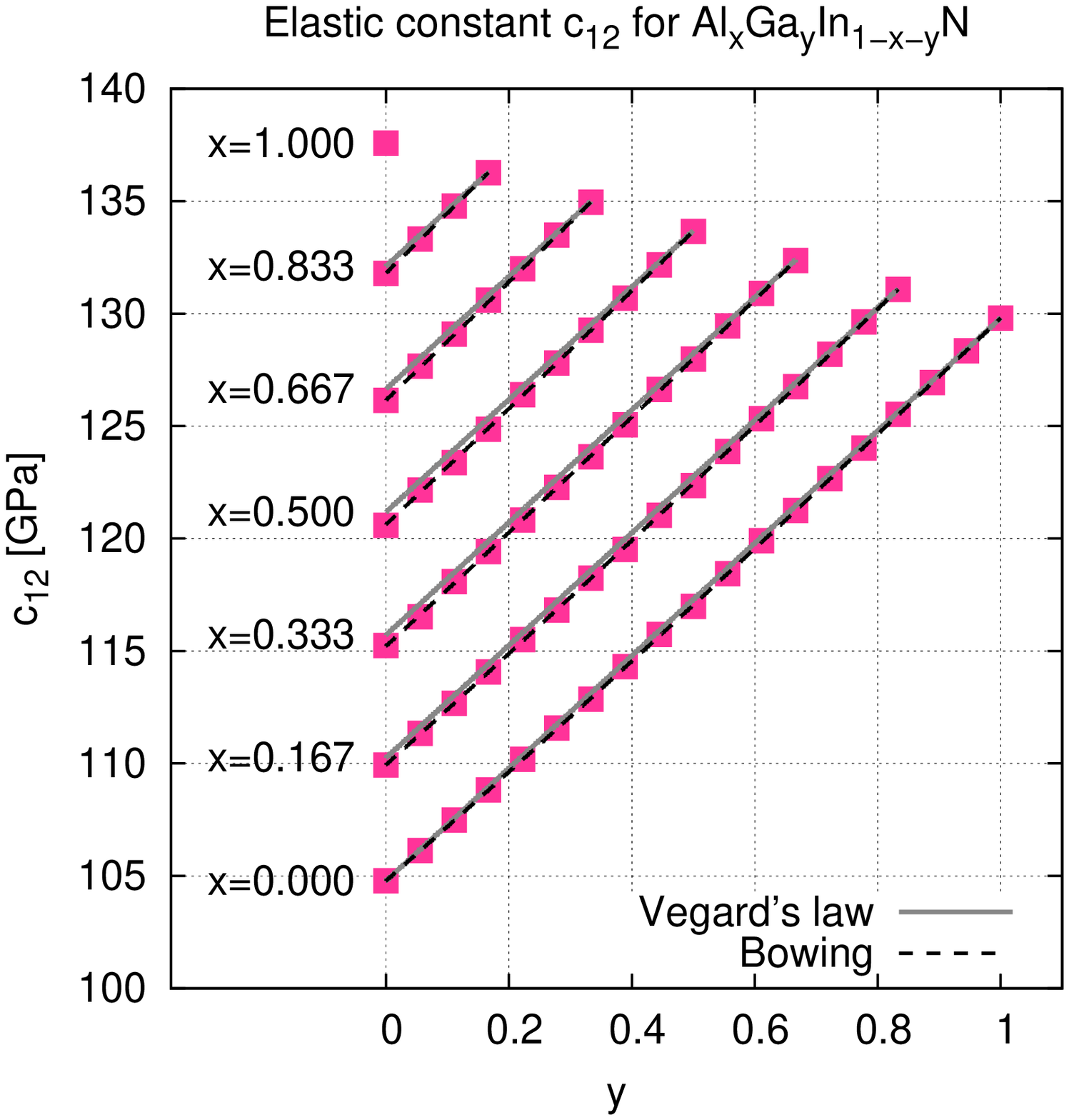} 
	\includegraphics[height=0.49\textwidth,angle=270]
			{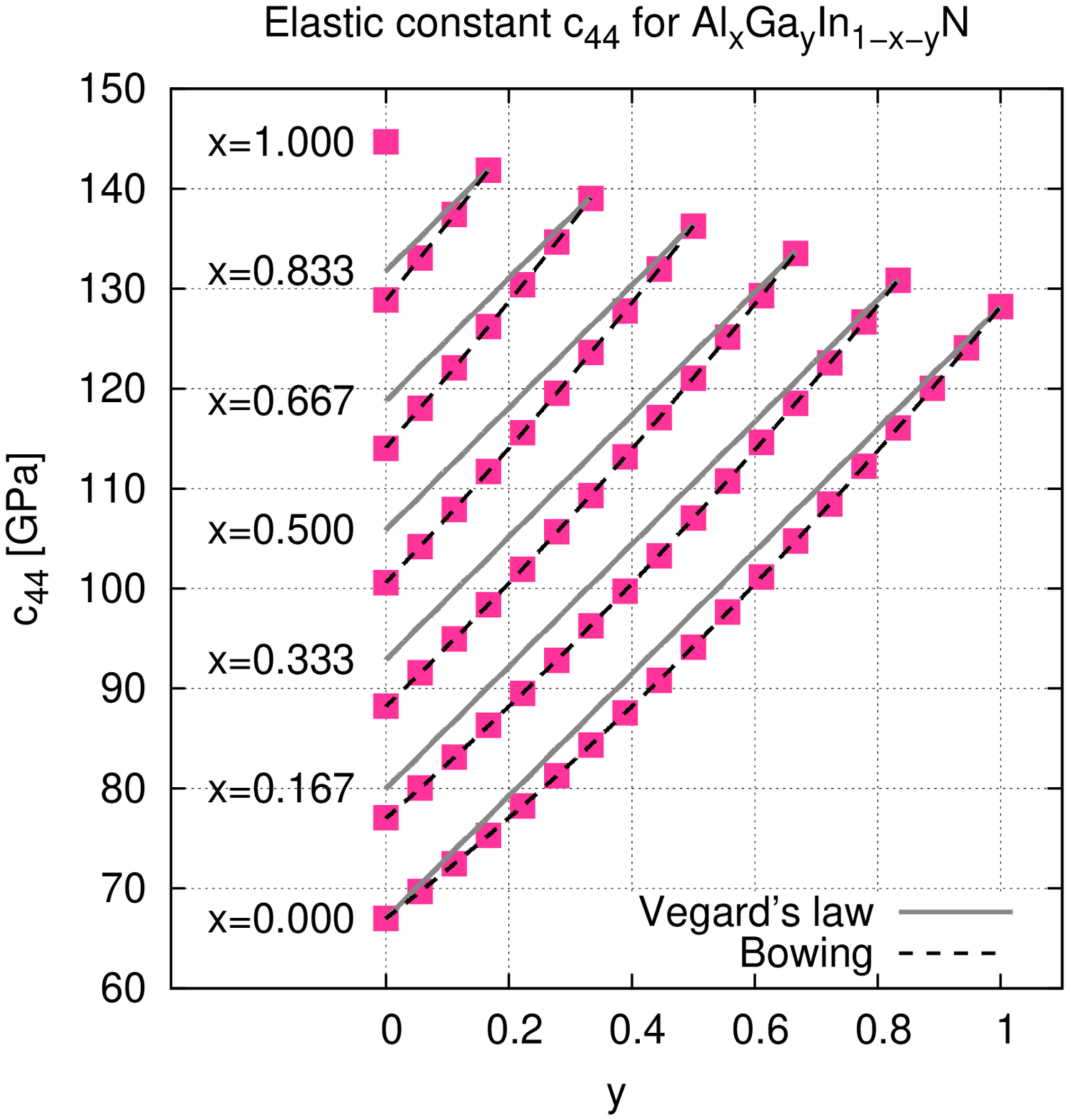} 
	\end{center}
	\caption{
 	 \label{Fig:ElasticConstants}
	   Elastic constants dependence on composition
	   for \AlGaInN alloys calculated on the basis of Keating VFF.
	}
\end{figure}

\begin{table}[ht!]
\caption{\label{Tab:ElasticConstantsAlloys}
Concentration dependence of the elastic constants  $c_{ij}$ 
for \AlGaInN alloys. Accuracy of linear Vegard-like model is  
compared with fits including additional bowing term $\Delta c_{ij}$.
All data in GPa. }
\vspace{0.2cm}
\begin{indented}
\item[] \begin{tabular}{l|l|cc}   
\rule{0pt}{3ex}
         & Result of fit for $c_{ij}$
         & \multicolumn{2}{c}{Max. difference:} \\
	 & 
	 & \multicolumn{2}{c}{[GPa]} \\    
\hline
$c_{11}$ & \myStrut Vegard's law      &        &       \\
         & $182.23 +135.79\,x +105.70\,y$ 
         & 6.2 & (2.6\%) \\
         & \myStrut Vegard's law + bowing term $ \Delta c_{11}$ 
		 &        &       \\
         & $\Delta c_{11}=-24.51\,x(1-x)+39.70\,x y -15.44\,y(1-y)$ 
	 & 0.3 & (0.1\%) \\
\hline                        
$c_{12}$ & \myStrut Vegard's law             &        &       \\
         & $104.78 +32.80\,x +25.02\,y$
         & 0.7 & (0.6\%) \\
         & \myStrut Vegard's law + bowing term $\Delta c_{12}$   
         &        &       \\
         & $\Delta c_{12}=-2.22\,x(1-x) +3.10\,x y -1.00\,y(1-y)$ 
         & 0.1 & (0.1\%) \\
\hline
$c_{44}$ & \myStrut Vegard's law             &        &       \\
         & $67.00 +77.71\,x +61.20\,y$
	 	 & 5.3 & (5\%) \\
	 	 & \myStrut Vegard's law + bowing term $\Delta c_{44}$   
	 	 &        &       \\
         & $\Delta c_{44}=-21.06\,x(1-x) +33.99\,x y -13.68\,y(1-y)$
         & 0.1 & (0.1\%) \\
\end{tabular}
\end{indented}
\end{table}

The literature indicates that indeed bowing in the alloy elastic constants
$c_{ij}$ should be expected. Chen and Sher in their book \cite{Chen1995} point
out that the value of $\Delta c_{ij}$ should be always negative, which is the
case in our studies. They argue that since the elastic properties of
semiconductors correlate with inverse power of lattice constant and lattice
constants for alloys follows Vegard's law, then the $c_{ij}$ dependence on
composition should be sublinear. They also perform simple analysis within the
framework of Keating VFF model showing that for simple ordered structures
sublinear bowing in bulk modulus is present.  Also our preliminary calculations
within virtual crystal DFT pseudopotential scheme (VCA DFT, sometimes
referenced as computational alchemy) predict a presence of the bowing term in
$c_{ij}(x,y)$ \cite{Lopuszynski2007a}.  However, one has to bear in mind that
VCA DFT works best for alloys, when lattice mismatch of constituents is low.
Larger mismatch, (as in the case of InN with both AlN and GaN) can introduce
considerable inaccuracy to this model.

To summarize this section, we have shown that Keating VFF model predicts
quadratic dependence of the elastic constants $c_{ij}$ on \AlGaInN alloy's
composition. This effect is in agreement with the previous literature studies
\cite{Chen1995,Lopuszynski2007a}.  We believe that, even thought the effect is
not very large, the awareness of it could improve description and modelling of
devices based on low dimensional nitride structures, such as quantum dots and
quantum wells.  When the elastic constants of more common wurtzite phase of
\AlGaInN are needed, one can obtain desired dependencies using Martin
transformation \cite{Martin1972} to the data gathered in table
\ref{Tab:ElasticConstantsAlloys}.  It would be also interesting to compare
presented results with other modelling approaches, since Keating VFF is a
simple tool and does not capture many effects.  The accurate experimental
studies would be also of a great value here.

\section{Influence of mixing rule \label{Sec:MixingInfluence}
         and finite supercell size}
In this section we give a brief overview of two more technical aspects of
presented calculations. First, we examine to what extent the results of our VFF
alloy simulations are sensitive to the selection of mixing rule. Second, we
analyze how effects of finite supercell size influence the structural
properties.     

As already mentioned in section \ref{Sec:KeatingModelParam}, following
e.g. \cite{Mattila1999,Takayama2000,Takayama2001} and many other works,
we have used arithmetic mixing rule to interpolate between three-body
$\beta$ constants of base materials, i.e.,
\begin{equation}
\label{Eq:ArithMixing}
 \beta_{\rm{Al},\rm{N},\rm{Ga}}= 
	\frac{  \beta_{\rm{Al},\rm{N},\rm{Al}} + 
                \beta_{\rm{Ga},\rm{N},\rm{Ga}} }{2}
\end{equation}  
and so on. Obviously this is not the only option.
Therefore, important question arises - to what extent this choice influences
the results presented so far? In order to check this,  we carried out a set of 
simulations for geometric mixing, used e.g. by Schabel and Martins
\cite{Schabel1991} or Saito \etal \cite{Saito1999}. 
This means that instead of equation (\ref{Eq:ArithMixing})
we have:
\begin{equation}
 \beta_{\rm{Al},\rm{N},\rm{Ga}}= 
		\sqrt{\beta_{\rm{Al},\rm{N},\rm{Al}} 
                   \; \beta_{\rm{Ga},\rm{N},\rm{Ga}}}
\end{equation}  
and so on. Other Keating VFF parameters are left unchanged.

Sample results of this numerical experiment are presented in figure
\ref{Fig:CompareMixing}. Generally, it turns out that average lengths remain
virtually unaffected by the type of mixing.  On the left panel of figure
\ref{Fig:CompareMixing}, where average In-N  distance is presented, one can see
that curves for arithmetic and geometric mixing almost cover each other. This
is because the difference there does not exceed 0.05 \%. The behaviour is
similar for both average nearest neighbours and next nearest neighbours
distances.  Small differences could be observed in the results for elastic
constants.  Maximum deviations were: $\Delta c_{11} = 1.4\; \rm{GPa} \;
(0.6\%)$, $\Delta c_{12}=0.5 \; \rm{GPa} \; (0.4\%)$, $\Delta c_{44}=1.4 \;
\rm{GPa} \; (1.3\%)$.  It turns out that $c_{11}$ and $c_{44}$ are always
larger in arithmetic than in geometric mixing model, whereas for $c_{12}$, it
is the other way round. Graphical comparison of $c_{44}$ dependence on
concentration in both approaches is depicted in the right panel of figure
\ref{Fig:CompareMixing}.  Even though results for elastic properties depend on
the type of mixing used, differences are really very small, strongly suggesting
that both ways of mixing constants $\beta$ lead to equally valuable
predictions. Detailed verification which rule works better would be rather
difficult, requiring very accurate experimental investigation. 

\begin{figure}[!ht]
	\includegraphics[height=0.45\textwidth,angle=270]
			{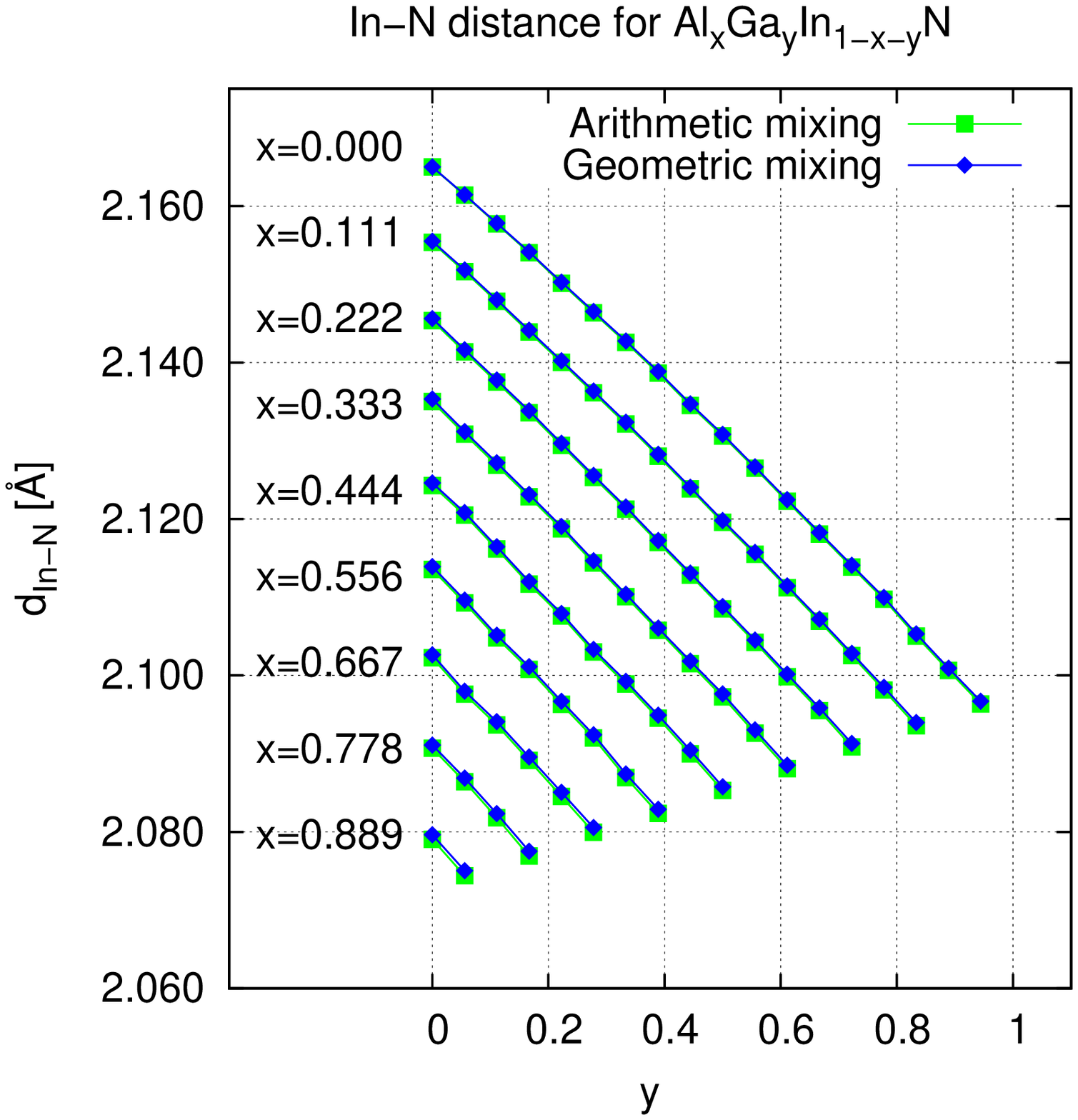}
	\includegraphics[height=0.45\textwidth,angle=270]
			{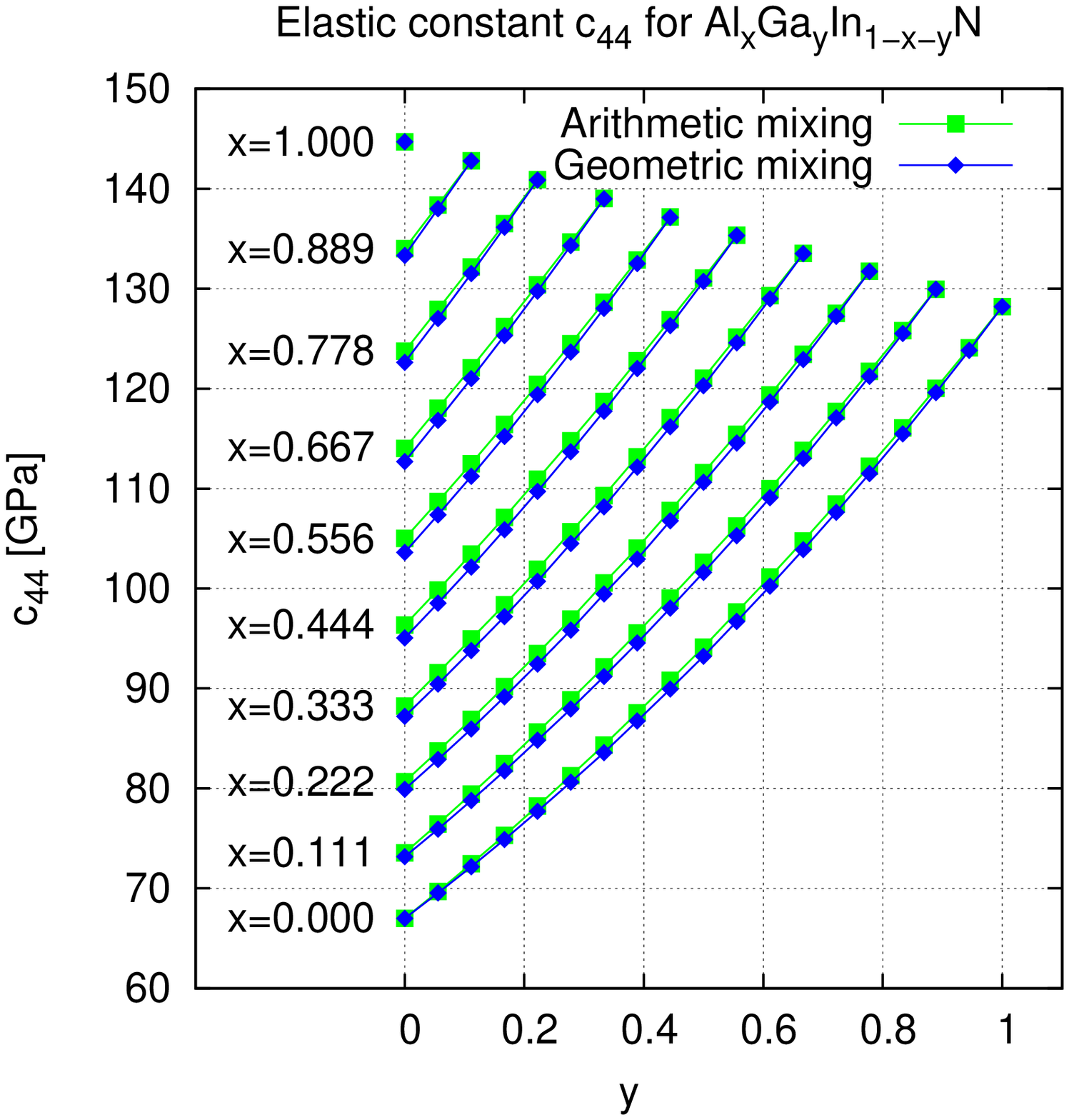} 
\caption{\label{Fig:CompareMixing} Comparison of calculations results
using VFF model with different mixing rules - 
arithmetic and geometric. Left panel presents $d_{\rm{In-N}}(x,y)$, 
the results for arithmetic and geometric mixing are so close to each other that
it is impossible to distinguish them in the figure. The 
right panel presents similar comparison for the elastic constant $c_{44}(x,y)$.}
\end{figure}

Another interesting technical aspect of presented simulations is verification
to what extent the finite cell size influences obtained results.  Particularly,
one may wonder if results obtained on the basis of two $3 \times 3 \times 3$
supercells per concentration in DFT LDA calculations (section
\ref{Sec:StrucProps}) reproduce sufficiently well the properties of random
alloy.  To verify this, we computed geometries of the same small cells used for
DFT LDA calculations by means of VFF model.  The comparison of data obtained
for this small cells with our VFF results for $18 \times 18 \times 18$ (46656
atoms) is presented in figure \ref{Fig:CompareFiniteSize}.  One can see that on
the sample diagram of nearest neighbour distance $d_{\rm{Al-N}}$, even though
differences are present, they have the form of typical statistical noise
preserving the same linear trends as it has been found for large supercells.
The same behaviour was observed for other pairs of the nearest neighbours.  The
analysis of the next nearest neighbour distances reveals even better agreement.
This is because the second coordination shell in zinc-blende (or ideal
wurtzite) contains 12 atoms, whereas the first consists of only 4 nearest
neighbours. This increases the statistics and leads to smaller error for the
next neighbours.  On the basis of presented comparison, one can conclude that
$3 \times 3 \times 3$ supercells reproduce correctly the trends observed in the
structural properties for much larger systems.  

\begin{figure}[!ht]
	\includegraphics[height=0.45\textwidth,angle=270]
			{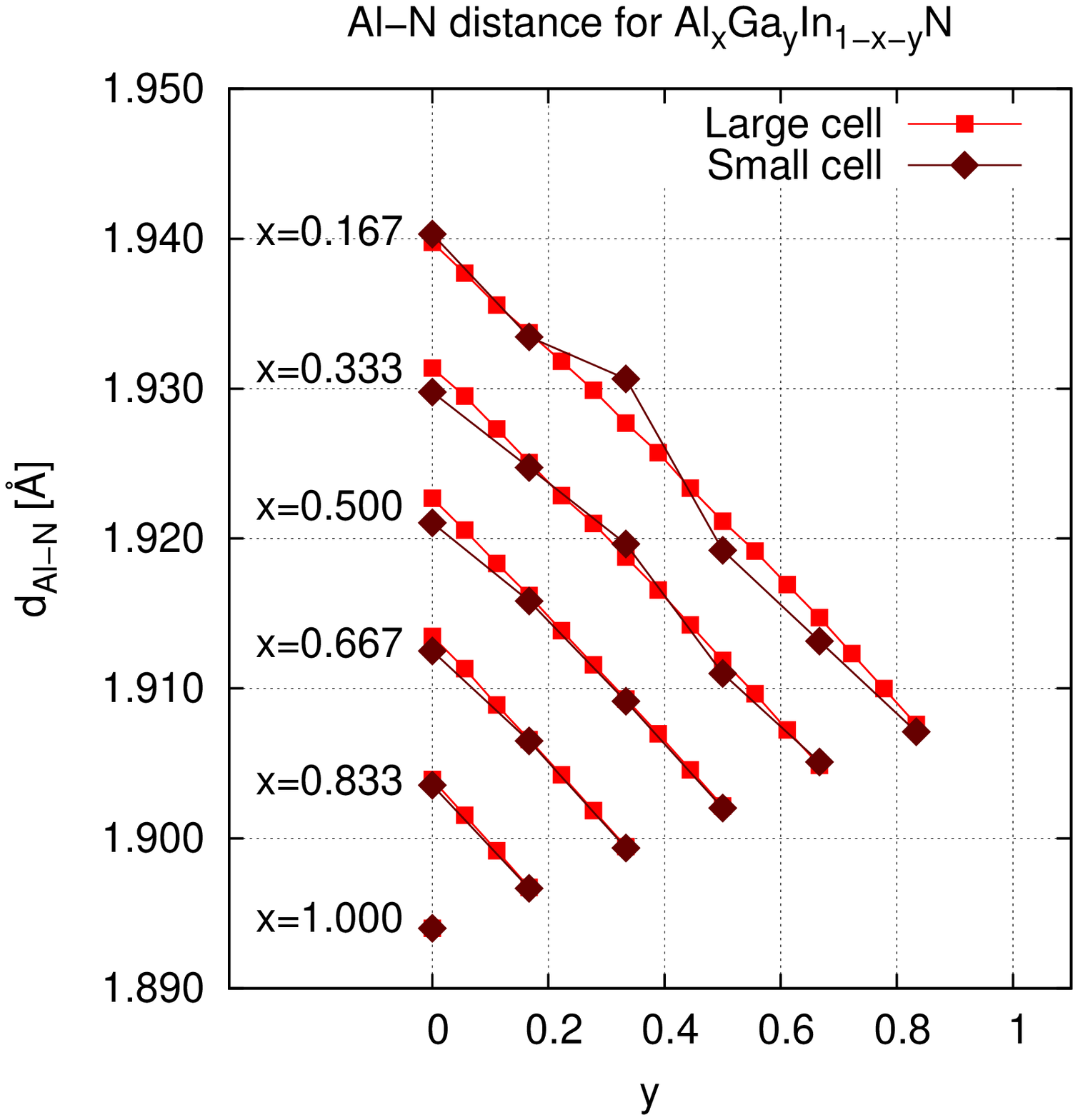}
	\includegraphics[height=0.45\textwidth,angle=270]
			{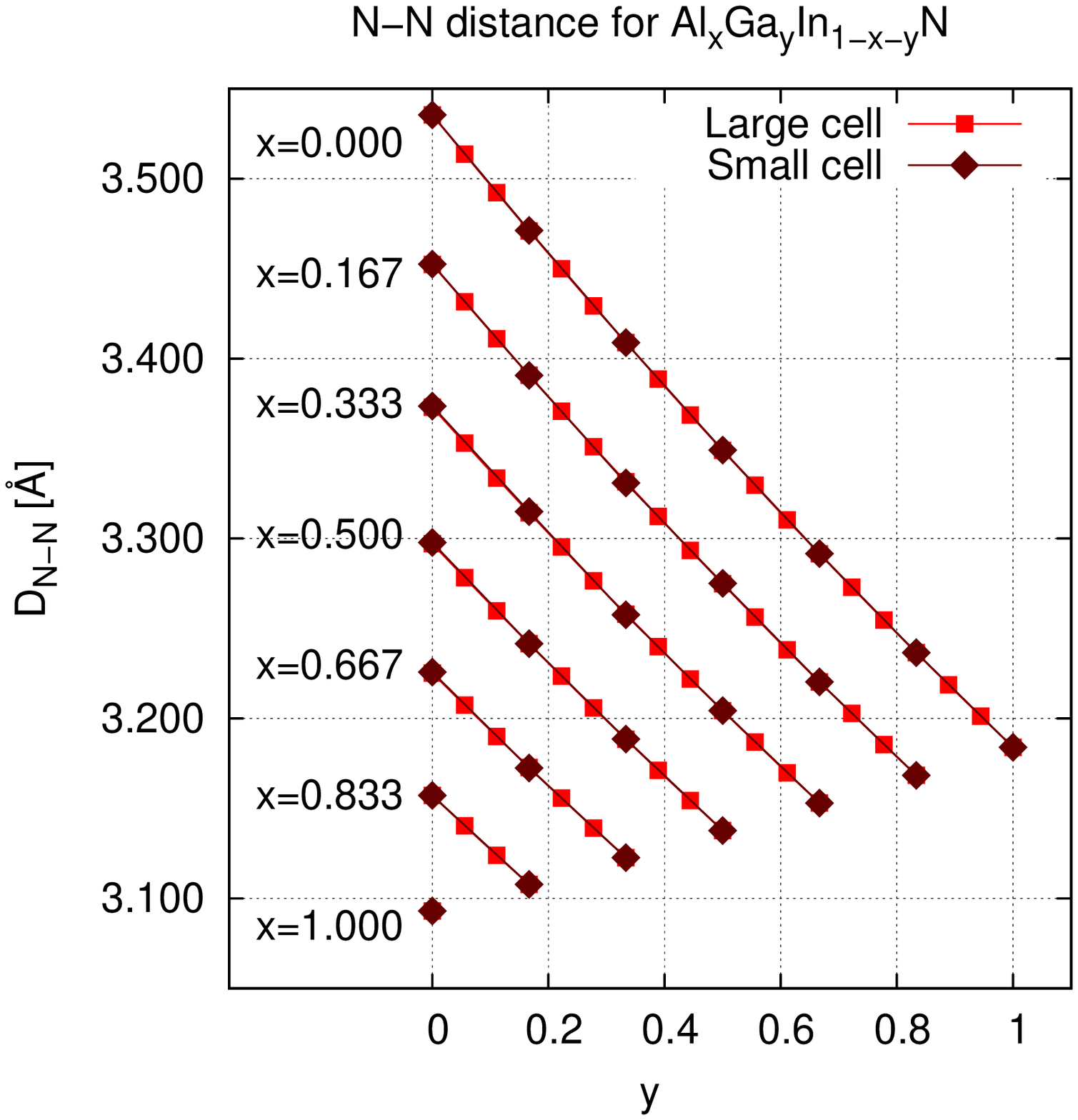}
\caption{\label{Fig:CompareFiniteSize} Comparison of calculations results using
VFF model with different cell sizes. Small cell size corresponds to data
gathered from two different $ 3 \times  3 \times  3$ cubic cells (216 atoms
each), large cell corresponds to $18 \times 18 \times 18$ cubic cell (46656
atoms).  Right panel compares nearest neighbour distance $d_{\rm{Al-N}}(x,y)$
and left panel presents next nearest neighbour distance $D_{N-N}(x,y)$.  The
latter results for both considered sizes are so close that they are impossible
to distinguish in the picture.} 
\end{figure}

\section{Summary \label{Sec:Summary}}
In this work we have presented computational study of structural and elastic
properties of zinc-blende quaternary \AlGaInN alloys over the whole
concentration range.  Our main computational tool was Keating VFF model. We
have started with presenting new parametrization of this model based on
state-of-the- art quantum-mechanical calculations within DFT formalism.  Then
we have shown the VFF results for lattice constant and distributions of the
nearest neighbours and the next nearest neighbour distances.  We have compared
these predictions with accurate DFT LDA calculations for supercells of moderate
size.  It has turned out that the agreement is reasonable, which shows that
simple nearest-neighbour interaction approximation made in Keating VFF
sufficiently well captures the most important aspects of more accurate DFT
picture.  Then we have also used VFF model to examine the elastic constants,
concluding that the composition dependence of $c_{ij}$  exhibits deviation from
Vegard-like model in form of sublinear bowing.  This is in accordance with
suggestions already made in the literature \cite{Chen1995,Lopuszynski2007a}.
We have also presented accurate quadratic function fits, which very well
approximate the dependence of $c_{ij}$ on composition including aforementioned
bowing effect.  This could be used to improve continuous models of
nanostructures.  Finally, we have examined the influence of mixing rules on VFF
results.  It has turned out that structural properties remain virtually
unaffected when one uses geometric mixing instead of arithmetic one.  The
effect on elastic constants is larger, however still much lower than typical
experimental error.  

\ack
This research was supported by the European Union within European Regional 
Development Fund, through grant Innovative Economy (POIG.01.01.02-00-008/08).

\bibliographystyle{iopart-num}
\bibliography{bibliography}

\end{document}